\RequirePackage{fix-cm}
\documentclass[natbib]{svjour3}         
\smartqed  

\usepackage[normalem]{ulem}

\newcount\listnorom
\newcommand\listromanDE{\global\advance \listnorom by 1
{\lowercase\expandafter{(\romannumeral\listnorom)}\ }}

\newcount\listnumber
\newcommand\listDE{\global\advance \listnumber by 1
{\lowercase\expandafter{(\number\listnumber)}\ }}
\newcommand\newlistDE{\listnumber=0}

\newcount\Inum
\newcount\IInum
\newcount\IIInum
\newcount\IVnum
\def\I{\global\multiply\IInum by 0 \global\multiply\IIInum by 0
            \global\multiply\IVnum by 0 \global\advance \Inum by 1
            {\the\Inum. }}
\def\II{\global\multiply\IIInum by 0\global\multiply\IVnum by 0
       \global\advance \IInum by 1 {\the\Inum.\the\IInum. }}
\def\III{\global\multiply\IVnum by 0\global\advance \IIInum by 1
            {\the\Inum.\the\IInum.\the\IIInum. }}
\def\IV{\global\advance \IVnum by 1
            {\the\IVnum. }}
\newcommand{\West}{Westerlund~1}

\usepackage{graphicx}
\usepackage{epsf}
\usepackage{amssymb}
\usepackage{amsmath}
\usepackage{placeins}
\usepackage{color}
\usepackage{multirow}

 \journalname{SSRv}
\newcommand{\MB}{Maxwell-Boltzmann}
\newcommand{\Msun}{\mbox{$M_{\odot}\;$}}
\newcommand{\Rsk}{R_\mathrm{sh}}
\newcommand{\usk}{u_\mathrm{sh}}
\newcommand{\Esn}{E_\mathrm{SN}}
\newcommand{\CasA}{Cassiopeia~A}
\newcommand{\SNRJ}{SNR RX J1713.7-3946}

\newcommand{\Jcr}{J_\mathrm{cr}}
\newcommand{\Bturb}{B_\mathrm{turb}}
\newcommand{\Qesc}{Q_\mathrm{esc}}
\newcommand{\gamZ}{\gamma_0}

\newcommand{\rgz}{r_{g0}}

\newcommand{\levy}{{\it L\'evy}}
\newcommand{\mfp}{mean free path}
\newcommand{\zFEB}{z_\mathrm{FEB}}
\newcommand{\FEB}{free escape boundary}
\newcommand{\Mdot}{\dot{\rm{M}}}
\newcommand{\Mej}{M_\mathrm{ej}}
\newcommand{\Emax}{E_\mathrm{max}}
\newcommand{\kmps}{km s$^{-1}$}
\newcommand{\degg}{^\circ}
\newcommand{\xx}[1]{\!\times\!10^{#1}}
\newcommand{\ISM}{interstellar medium}

\newcommand{\be}{\begin{equation}}
\newcommand{\ee}{\end{equation}}
\newcommand{\beq}{\begin{eqnarray}}
\newcommand{\eeq}{\end{eqnarray}}
\newcommand\subsun[1]{{$_{\normalsize\odot}$}}
\newcommand{\ergs}{erg~s$^{-1}$}   
\newcommand{\kms}{~km~s$^{-1}$}

\newcommand{\mc}{Monte Carlo}

\newcommand{\FoFSA}{first-order Fermi shock acceleration}
\newcommand{\FFoFSA}{First-order Fermi shock acceleration}
\newcommand{\Facc}{Fermi acceleration}
\newcommand{\DSA}{diffusive shock acceleration}
\newcommand{\CSM}{circumstellar medium}
\newcommand{\MFA}{magnetic field amplification}

\definecolor{pink}{rgb}{0.91, 0.67, 0.81}
\definecolor{violet}{rgb}{0.93, 0.51, 0.93}

\def\apjl{ApJ}%
\def\apjs{ApJS}%
\def\ao{Appl.~Opt.}%
\def\aap{A\&A}%
\def\lsim{\;\raise0.3ex\hbox{$<$\kern-0.75em\raise-1.1ex\hbox{$\sim$}}\;}
\def\gsim{\;\raise0.3ex\hbox{$>$\kern-0.75em\raise-1.1ex\hbox{$\sim$}}\;}
\def\alf{Alfv\'en }

\def\diff{\rm ~cm^2~s^{-1}}

\def\lsim{\;\raise0.3ex\hbox{$<$\kern-0.75em\raise-1.1ex\hbox{$\sim$}}\;}
\def\gsim{\;\raise0.3ex\hbox{$>$\kern-0.75em\raise-1.1ex\hbox{$\sim$}}\;}

\def\kms{\rm ~km~s^{-1}}

\def\diff{\rm ~cm^2~s^{-1}}
\def \kms {\rm ~km~s^{-1}}

\def\aap{A\&A}                   
\def\apjs{ApJS}                  
\def\apjl{ApJ}                   

\def\lsim{\;\raise0.3ex\hbox{$<$\kern-0.75em\raise-1.1ex\hbox{$\sim$}}\;}
\def\gsim{\;\raise0.3ex\hbox{$>$\kern-0.75em\raise-1.1ex\hbox{$\sim$}}\;}

\definecolor{purple}{rgb}{0.63, 0.36, 0.94}

\def\kms{\rm ~km~s^{-1}}

\def\aap{A\&A}

\def\apjl{ApJ Letters}

\def\apjl{ApJ}%
\def\apjs{ApJS}%
\def\ao{Appl.~Opt.}%
\def\aap{A\&A}%
\newcommand{\TP}{test-particle}
\newcommand{\NL}{nonlinear}
\newcommand{\SCly}{self-consistently}
\newcommand{\SC}{self-consistent}

\newcommand{\Ss}{self-similar}
\newcommand{\gamray}{$\gamma$-ray}
\newcommand{\gamrays}{$\gamma$-rays}
\newcommand{\nonrel}{non-relativistic}
\newcommand{\rel}{relativistic}

\newcommand{\transrel}{trans-rel\-a\-tiv\-is\-tic}

\newcommand{\ultrarel}{ul\-tra-rel\-a\-tiv\-is\-tic}

\newcommand{\Lor}{Lorentz}
\newcommand{\RH}{Rankine-Hugoniot}

\newcommand{\NEI}{non-equilibrium ionization}
\newcommand{\syn}{synchrotron}
\newcommand{\synch}{synchrotron}
\newcommand{\pion}{pion-decay}
\newcommand{\IC}{inverse-Compton}
\newcommand{\brem}{bremsstrahlung} 
\newcommand{\brems}{bremsstrahlung} 

\begin{document}
\title{Cosmic ray production in supernovae
}
\titlerunning{Cosmic ray production in supernovae}

\author{A.M.~Bykov \and D.C.~Ellison \and A. Marcowith \and S.M. Osipov}

\authorrunning{A.M.~Bykov et al.} 

\institute{A.M.~Bykov \at
           Ioffe Institute, 194021, St. Petersburg, Russia; ambykov@yahoo.com; \and  D.C.~Ellison \at North Carolina State University, Department of Physics, Raleigh, NC 27695-8202, USA; don\_ellison@ncsu.edu \and A.~Marcowith  \at Laboratoire Univers et Particules de Montpellier CNRS/Universite de Montpellier,
Place E. Bataillon, 34095 Montpellier, France; Alexandre.Marcowith@umontpellier.fr \and S.M.~Osipov \at
Ioffe Institute, 194021, St. Petersburg, Russia; osm2004@mail.ru}

\date{Received: 4 December 2017 / Accepted in Space Science Reviews: 24 January 2018 \\ \\
Supernovae \\ 
Edited by Andrei Bykov, Roger Chevalier, John Raymond, Friedrich-Karl Thielemann, Maurizio Falanga and Rudolf von Steiger}

\maketitle

\begin{abstract}
We give a brief review of the origin and
acceleration of cosmic rays
(CRs),  emphasizing the production of CRs at different stages of
supernova evolution by the \FoFSA\ mechanism.
We suggest that supernovae with
\transrel\ outflows, despite being rather rare, may accelerate CRs
to energies above 10$^{18}$\,eV over the first year of their
evolution.
Supernovae in young compact clusters of massive stars, and
interaction powered superluminous supernovae, may accelerate CRs well
above the PeV regime. We discuss the acceleration of the bulk of the
galactic CRs in isolated supernova remnants and re-acceleration of
escaped CRs by the multiple shocks present in superbubbles produced
by associations of OB stars.
The effects of magnetic field
amplification by CR driven instabilities, as well as superdiffusive CR
transport, are discussed  for nonthermal radiation produced by \NL\ shocks of all speeds including \transrel\ ones.
%
\end{abstract}


\section{Introduction} \label{sec:intro}
Based on early inconclusive evidence, \citet{BZ34} suggested that
cosmic rays (CRs) are produced by supernovae (SNe). Since then it
has become clear that  SNe are the most likely source of CRs, at
least for those with energies below the so-called ``knee" at $\sim
10^{15}$\,eV, for four main reasons.
First, and most important, SNe are the only known galactic source
with sufficient energy to power CRs
\citep[e.g.,][]{Ginzburg1964,Axford81,Berezinski90,Hillas2005,Reynolds08,Lingenfelter17}.
Even for supernova remnants (SNRs) the acceleration of CRs must be highly efficient.
Second, the source composition of the bulk of CR material is
primarily well-mixed \ISM\ (ISM) material (including those elements
preferentially locked in dust) with no more than  a $20\%$
contribution from fresh core-collapse SN ejecta material
\citep[i.e.,][]{MDE97,EDM97,ME1999,BinnsEtal2014}.
This requires a source that injects {\it old} (i.e., not freshly
synthesized) material from the entire galaxy including material from
low-mass stars that never explode.  No source other than SNRs has a
galactic filling factor large enough to do this.
Third, there is now a wealth of observational evidence from SNRs
showing non-thermal radiation emitted by \ultrarel\ electrons and
ions \citep[e.g.,][]{helder12,AckermannEtal_W44,
blasi2013,amato14,blandford14,Slane2015Err}. The {\it in situ}
production of CRs is observed in SNRs!
Fourth, the collisionless shocks associated with SNRs can utilize
the \FoFSA\ mechanism to simultaneously accelerate various ion
species with high efficiency \citep[e.g.,][]{AhnEtal2010}. Perhaps
even more important, for the same set of parameters, \FoFSA\
accelerates  electrons and ions with similar spectral shapes, as
observed in CRs and solar energetic particles
\citep[e.g.,][]{ER85,BoyleEtal2008}. No other mechanism as naturally
produces similar spectral shapes for leptons and hadrons.

The power required to maintain the observed energy density
of CRs (dominated by nuclei in the GeV range)
is
$\sim 2\xx{41}$\ergs\  \cite[e.g.,][]{Berezinski90}.
This is  estimated by assuming some CR propagation model with the CR escape length from the galaxy derived from the observed secondary to primary nuclei ratio
\cite[e.g.,][]{Ptuskin2012,ABPW2012}.
If in the Milky Way, there is one SN every 30--100\,yr with an
average kinetic energy of $2\xx{51}$\,erg, an efficiency of 10--30\%
is required to power CRs, i.e., $>10\%$ of the ram kinetic energy of
the SNR blast wave must be put into \rel\ particles of the GeV
energy regime.

If acceleration  efficiencies are this large, \NL\ feedback effects from CR
production will influence the shock dynamics and the resultant CR
spectral shape \citep[e.g.,][]{BE99}. Consequences for the dynamics
include the fact that the \TP\ relation between the postshock
temperature, $T_2$, and the shock speed, $u_0$, often assumed for
strong  shocks, will be modified, i.e., $(1/\mu m_H)(kT_2/u_0^2 \neq
3/16)$ \citep[e.g.,][]{DEB2000,HRD2000}. Here $\mu m_H$ is the mean
particle mass. Likewise,  the Sedov solutions relating the SNR age
to the shock radius and speed will also be modified
\citep[e.g.,][]{EPSBG2007}.

Recent detailed observations of the energy  spectrum, composition,
and  anisotropy of the angular distribution of CRs up to TeV
energies have been made by spacecraft and balloon instruments, e.g.,
{\sl ACE}, {\sl PAMELA}, {\sl AMS2}, {\sl CREAM}, {\sl SuperTIGER},
and {\sl Fermi}. For higher energies, ground-based experiments are
required and these include {\sl KASCADE-Grande}, {\sl TUNKA}, and
{\sl LOFAR} up to a few hundred PeV, while the {\sl Telescope Array}
and the {\sl Pierre Auger} observatories provided detailed CR
observations to EeV energies.
The {\sl LOFAR} array measures radio emission produced by \rel\
leptons created in high-energy CR air showers. This relatively new
technique has a large duty cycle and can determine the atmospheric
depth of shower maximum, and thus CR energy, with high resolution
\citep[][]{BuitinkEtal2016}.
The high statistics of the above mentioned measurements have allowed
a closer look into the spectral features which have long been used
to understand the sources of CRs, their propagation in the Galaxy
and beyond, and the assumed transition from a galactic source to
extragalactic sources of CRs above $\sim 10^{18}$\,eV
\citep[e.g.,][]{Bergman2007,Lipari2017}.

A particularly interesting feature has been the spectral hardening
of protons and helium at rigidities above $\sim 230$\,GV reported by
{\sl PAMELA} and {\sl AMS2}.\footnote{Rigidity is defined as $pc/Ze$
where $p$ is the particle momentum, $c$ is the speed of light, $e$
is the electronic charge, $Z$ is the total charge number of the
particle, and the units are volts.}
This hardening is  likely fairly broad and may extend beyond the
rigidity range of both {\sl PAMELA} and {\sl AMS2}.
The combined fit of {\sl AMS2} and {\sl
CREAM} data made by \citet{Lipari2017} indicates a very broad
feature that extends from 200 GeV to 2 TeV  with a spectral index
difference of $\sim 0.2$ \citep[see][]{Malkov2017}.

Other important features  in the all particle spectrum are the
so-called ``knee" at $10^{15-16}$\,eV and the ``ankle" in the EeV
range. The energy band  between the knee and the ankle is crucial
for understanding the transition between galactic and extragalactic
CRs \citep[e.g.,][]{Bergman2007,AloisioEtal2007}.
Regardless of the energy  range, and despite decades of portraying
the all particle spectrum observed at Earth as a simple power law,
improved statistics are beginning to show that irregularities and
substructures are present in the spectrum
\citep[e.g.,][]{ApelEtal2013,berezhnevea13}.

Recent high resolution measurements by the {\sl LOFAR} telescope
below $\sim 1$\,EeV \citep{BuitinkEtal2016} favor a CR composition
with a light-mass fraction (protons and helium nuclei) of about
80\%. A light-mass fraction this high is surprising and may show the
existence of an extragalactic component at energies between the knee
and the ankle, or a previously unsuspected light-mass galactic CR
component at these energies \citep[e.g.,][]{Thoudam2016}.

Another important  measurement is the anisotropy. This is below a
few percent for CR energies $\lsim 1$\, EeV.  The data thus far does
not allow clear conclusions at higher energies but recent
observations have shown a large-scale anisotropy with an amplitude
of $\sim 10^{-3}$, along with small-scale structure (angular size
$10\degg - 30\degg$) at  about the $10^{-4}$ level
\citep[i.e.,][]{HAWC_35ICRC}. It is still not possible to constrain
particular source populations, or specific nearby sources, from the
anisotropy observations.

The direct observation  of CRs can only be done at the Earth since
the charged CRs meander through the irregular galactic
magnetic field and virtually all directional information from the
source is lost.
Source information  can be obtained, however, by observing the
non-thermal radiation CRs produce in specific objects such as SNRs.
Multi-wavelength observations of SNRs give detailed information
on the CR  acceleration process. Diffuse galactic emission on the
other hand gives direct information on the distribution of CRs
throughout the galaxy.

Recent observations of non-thermal
radiation from young ($<1000$\,yr) and middle-aged
SNRs have proven that particles are accelerated to above 10
TeV by the shocks in these objects.
Synchrotron radiation from \rel\  leptons is
detected from radio to X-rays, and the high angular
resolution observations by {\sl VLA} and {\sl Chandra} of young
SNRs like Tycho's SNR, Cas A, SN1006, and RX J1713.7-3946
provide compelling evidence for \MFA\ (MFA) directly associated with the acceleration process at the remnant forward shock
\citep[e.g.,][]{Reynolds08,vink12,helder12}.

The Fermi acceleration
 process \citep[i.e.,][]{Fermi_PR49,Fermi_ApJ54}, and its
particular realization called diffusive shock acceleration (DSA)
\citep[][]{ALS77,Kry77,Bell78a,BO78}, provides
a plausible way to efficiently convert the
kinetic energy released in a supernova
shock to a wide energy spectrum of accelerated particles
\citep[e.g.,][]{be87,je91,MD2001,SchureEtal2012}.
In fact, \Facc\ can occur at any collisionless shock and there is direct observational confirmation of efficient particle acceleration at the
quasi-parallel Earth bow shock \citep[i.e.,][]{EMP90}.\footnote{We note that while the terms ``\Facc" and ``\DSA" are often used interchangeably there is an important difference when \rel\ shocks are discussed. Particle transport in \rel\ flows need not be diffusive. In fact, non-diffusive behavior in steep density gradients, as near the sharp subshock transition and at an upstream \FEB, is critically important for all aspects of particle acceleration and MFA. Henceforth we use \Facc\ to include all cases including those where diffusive behavior does dominate.}

The forward shocks in most galactic SNRs are observed to have speeds
below $\sim 10^4$\,\kmps. While these shocks are believed to be
capable of producing the bulk of galactic CRs to energies
approaching the knee via \Facc\ uncertainties remain.
It is still uncertain how  particles escape from the accelerator
without experiencing strong adiabatic losses, and the maximum  CR
energy  produced in observed SNRs does not extend into the knee.
Particle escape depends on the galactic environment at the late
stages of SNR evolution which may be in the warm ISM or in a
superbubble produced by clustered SNe
\citep[e.g.,][]{maclow_mccray88,heiles90,Cox2005}.
As for the maximum CR  energy a given shock can produce, $\Emax$,
\citet{LC83} estimated $\Emax$ for \Facc\ at the forward shock of an isolated SNR over its lifetime as
$\Emax \sim 10^5~Z~B_{-6}$ GeV, where $Z$ is the ion charge number and $B_{-6}$ is
the ISM magnetic field measured in $\mu$G \citep[see
also][]{Hillas2005}.

The \gamray\
observations of SNRs W44 and IC~443 provide clear evidence for pion
production in a SNR due to TeV CR proton interactions
\citep{GiulianiEtal2011,AckermannEtal_W44}. The evidence for PeV CRs is less certain.
All of the SNRs identified so far show breaks in their \gamray\
spectra well below 100 TeV \citep[e.g.,][]{Funk2015}. These remnants
have been observed with the currently operating ground-based
Cherenkov telescopes {\sl H.E.S.S}, {\sl MAGIC} and {\sl VERITAS}.

We note that there is an  unidentified diffuse
H.E.S.S. source in the Galactic Center (GC) region
\citep[i.e.,][]{HESS2016} with  no clear spectral break or
cut-off until tens of TeV and this may well be a PeV CR accelerator.
There are a number of possible interpretations of this \gamray\
source. The Pevatron can be associated with the massive black hole
Sgr A$^{\star}$ \citep[e.g.,][]{Aharonian2005,Fujita2017,GuoEtal2017}
which may produce a blast wave after a tidal disruption event to
accelerate particles \citep[e.g.,][]{LiuEtal2016}.
An alternative explanation is that a supernova exploded  close to
the GC and interacted with either the fast wind from the GC or with
a wind from its parent compact cluster of massive stars. We will
discuss such a ``colliding shock flow system"  later in this paper
and give estimates of the maximum CR energies achievable at very
different evolution stages and for different types of SNe, as well
as the possible role superbubbles play
\citep[][]{Bykov2001,parizot04,
Ferrand2010,AckermannSB2011,Bykov2014,ohm16}. 

The review lay out is as follows. In \S \ref{S:OBN} we detail the observational status of particle acceleration in SNe and in SNRs.
In \S \ref{S:PEV} we discuss CR acceleration beyond the PeV regime by \transrel\ SNe or in superluminous SNe.
In \S \ref{s:hydro} we describe the different ingredients relevant for the investigation of particle acceleration in fast SN shocks as well as the expected radiation
spectra produced at fast shocks.
Section~\ref{S:OTH}  discusses other
galactic sources of PeV-EeV CRs: superbubbles and  SNe in clusters.
Section~\ref{S:OBS} addresses future observational
facilities which should result in a substantial improvement in our understanding of CR
acceleration in SNe.

\section{Observational status}\label{S:OBN}
In this section we discuss some observations of SNe.
In Section~\ref{sec:radio} we consider mainly radio observations of extragalactic SNe where the SN explosion and its immediate aftermath can be observed.
In Section~\ref{S:SNR} we consider evidence for MFA from galactic SNRs where detailed observations can be made from radio to \gamrays.

\subsection{Radio supernov\ae} \label{sec:radio}
Today about 200 extragalactic SNe have been detected in radio, but only seven are sufficiently close to show well-resolved light curves at radio wavebands. These are SN 1979C (SN IIL), SN 1986J (SN IIn), SN 1987A (SN IIpec), SN 1993J (SN IIb), SN1996cr (SN IIn), SN 2008iz,
and SN 2011dh (SN IIb) \citep[][and references therein]{Bartel17}, where the type of the core-collapse SN has been added in parenthesis when available. %
Among the objects in this list we discard SN 1987A and SN 1996cr. The former shows peculiar behavior difficult to account for in the following simplified models and the latter does not have enough data to constrain the shock dynamics and magnetic fields. Considering type Ib/c SNe, we have selected the following objects
on the basis of the amount of radio data available: SN 1983N, SN 1994I, SN 2003L \citep{Weiler86, Soderberg05}, and discuss the case of SN 2009bb since it is a relativistic Ibc SN \citep{Soderberg2010} (see also \S \ref{sec:trans}). 

Radio emission  is due to synchrotron radiation from electrons
accelerated by turbulent magnetic fields which develop in a region
including both reverse and forward shocks. However, the exact nature
of these turbulent magnetic fields, and the exact region of
acceleration, remain elusive \citep{Bjornsson17}.  There are
different locations where such turbulent magnetic fields can
develop: at the contact discontinuity where Rayleigh-Taylor
instabilities develop with fingers of the ejecta stretching into the
shocked \CSM\ (CSM), and/or at the forward or reverse shock where
existing turbulence can be amplified and the acceleration of CRs in
high-speed shocks can drive new MFA. Modeling of the radio
lightcurves in different wavebands requires accounting for a number
of processes.
These include synchrotron self-absorption (SSA), free-free absorption by the ambient thermal plasma (an internal effect) or by CSM matter (an external effect), and possibly plasma processes like the Razin-Tsytovich effect  \citep{Fransson98}. 

Radio observations are important for understanding many aspects of particle acceleration. First, the spectral turnover produced by SSA leads to an estimate of the magnetic field intensity of the synchrotron
emitting zone. Second, the synchrotron spectral index provides a constraint on the electron distribution function and then on the acceleration process. Third, radio images are used to derive the SN shell dynamics, time evolution of the shock radius and its velocity, both quantities mandatory for any microphysical calculation of particle acceleration efficiency \citep[e.g.,][]{Tatischeff09}. The shell radius and speed can also be compared to a self-similar
expansion model \citep[e.g.,][]{ch82}. 

The properties of the radio emission depend on the SN type: type
Ib/c SNe show steep spectral indices ($\alpha > 1$, with a radio
flux scaling as $S_{\nu} \propto \nu^{-\alpha}$), and have similar
flux peaking before optical maximum at a wavelength around 6~cm,
while other type II SNe show flatter spectra ($\alpha < 1$) with a
wider range of radio luminosities usually peaking at 6~cm
significantly after optical maximum  \citep[e.g.,][]{Weiler02}.
\cite{Bjornsson17} suggested that the radio emission in type Ib/c
SNe comes from a narrow region in the vicinity of the forward shock,
while the radio emission region of type IIb SNe (e.g. SN 1993J) is
wider due to the effect of the Rayleigh–Taylor instability.

In tables 1 and 2 we summarize the main properties of the above  SNe from the available data deduced from VLBI observations.
In addition to $\alpha$, we use the following notation.
The magnetic field strength is characterized by its amplitude $B_0$ at a reference time $t_0$ to be specified and the spectral index $n>0$ of its time variation: $B(t) = B_0 (t/t_0)^{-n}$. The VLBI shell radius is characterized by its radius at $t_0$, $R_0$ and by the index $m >0$ of its time variation: $R_{\rm sh} = R_0 (t/t_0)^{m}$. The shell velocity can then be deduced by $V_{\rm sh} = dR_{\rm sh}/dt$. Notice that the time evolution of the magnetic field is proportional to the energy density
in the shock as described by \cite{Marti11}
\begin{equation}\label{Eq:SSM}
B \propto \rho V_{\rm sh}^2 \propto t^{m{2-s \over 2} -1} \ .
\end{equation}
Here the mass density profile of the pre-SN stellar wind is assumed to be $\rho \propto r^{-s}$.
In the case of a shock propagating in a wind profile with a constant mass loss rate, that is with $s=2$, we find  from Eq.~(\ref{Eq:SSM}) an index for the magnetic field dependence of $n=1$.

\begin{table*} 
\begin{tabular}{|c| c| c| c| c| c|}
\hline 
Name  & $\alpha$ & $t_0(\rm{days})$ & $([B_0(\rm{G})], n)$ & $(R_0(\rm{cm}), m)$ \\
\hline
SN 1979C & $0.74^{+0.05}_{-0.08}$ & 5 & $([20-30], -1.00)$ & $(8.68(e14), 0.91\pm0.09)$ \\
\hline
SN 1986J & $0.67^{+0.04}_{-0.08}$ &5 & $([30-50], -1.00)$ & $(3.18(e15), 0.69\pm0.03)$ \\
\hline
SN 1993J & 1.00 & 10 & $([25.5], -0.93\pm0.08)$ & $(1.9(e15), 1.00)$  \\
\dots
                                        & 0.90 & 100 & $([2.4 \pm 1.0], -1.16\pm0.20)$ & $(1.6(e16), 0.829\pm0.005)$ \\
\hline
SN 2008iz & 1.00 & 100 & $([0.2-1.5], -1.00)$ & $(2.1(e16), 0.86\pm0.02)$ \\
\hline
SN 2011dh & 1.15 & 4 & $([5.9], -1.00\pm0.12)$ & $(5.0(e14), 1.14\pm0.24)$ \\
\dots
    & 0.95 & 15 & $([1.1], -1.00)$ & $(3.1(e15), 0.87\pm0.07)$  \\
\hline
\end{tabular}
\caption{Magnetic field and shock radius evolution in a set of type II SNe. The spectral index $\alpha$ corresponds to the optically thin synchrotron spectrum. Estimations for SN 1979C and SN 1986J are based on a SSA model. Magnetic field strengths are derived at day 5 and
are compatible with the strength of equipartition magnetic field given in \citet{Marti11}.
The value for $n$ is obtained from the solution of propagation in a wind with constant mass loss rate with $s=2$ (see Eq.~\ref{Eq:SSM}). SN 1993J: \citet{Fransson98} use an expansion index $m=0.74$ at $t > 100$ days,
\citet{Marti11} present a long term radio survey where the expansion law index at frequencies above 1.7 GHz varies from $m \le 0.925 \pm 0.016$ before $t_{\rm br}= 360\pm 50$ days to $m \ge 0.87\pm 0.02$ after. SN2008iz: \citet{Kimani16} derive the equipartition magnetic field strength, the lower and upper solutions depend if protons are not or are accounted in the estimation. References: \citet{Weiler91, Marcaide09, Marti11} for SN 1979C,
\citet{Weiler90, Bietenholz10, Marti11} for SN 1986J,  \citet{Fransson98, Tatischeff09} for SN 1993J,
\citet{Kimani16} for SN 2008iz, \citet{Horesh13, Krauss12, Yadav16} for SN 2011dh.}
\end{table*}

\begin{table*} 
\begin{tabular}{|c| c| c| c| c| c|}
\hline 
Name  & $\alpha$ & $t_0(\rm{days})$ & $(B_0(\rm{G}), n)$ & $(R_0(\rm{cm}), m)$ \\
\hline
SN 1983N  & $1.03\pm0.06 $ & 13 & $(3.6, -1.00?)$ & $(2.3(e15), 0.81)$ \\
\hline
SN 1994I & $1.22 $ & 10.125 & $(2.3, -)$ & $(2.39(e15), -)$ \\
\hline
SN 2003L & $1.1$ &10 & $(4.5, -1.00)$ & $(4.30(e15), 0.96)$ \\
\hline
SN 2009bb & $1.0$ &20 & $(0.6, -1.00)$ & $(3.20(e16), 1.00)$ \\
\hline
\end{tabular}
\caption{Magnetic field and shock radius evolution in a set of type Ib/Ic SNe. In the case of SN 1983N, the magnetic field strength dependence with time is assumed, \citet{Slysh92} only derives
estimates of an upstream magnetic field strength of 0.9 G which is multiplied by 4 here. In the case of SN 1994I, the fitted parameters of the model derived by \citet{Alexander15} do not show
a simple power-law time dependence for both $B$ and $R$.
References: \citet{Weiler86, Slysh92} for SN 1983N,  \citet{Weiler11, Alexander15} for SN 1994I, \citet{Soderberg05} for SN 2003L,
\citet{Soderberg2010, Chakraborti11} for SN 2009bb.}
\end{table*}

What is obvious from these tables is that the magnetic field quite early
in the SN evolution is, in general, far in excess of critical magnetic
field strengths obtained by balancing magnetic and wind
kinetic energies \citep[e.g.,][]{ud-Doula2002}.
The critical magnetic field  is given by $B^2_{\rm cr}/8\pi  = \rho v^2_{\rm w}/ 2$, where $v_{\rm w}$  is the
stellar wind speed of the massive star. A general value in Gauss is
\begin{equation}
\label{Eq:BWI}
B_{\rm  w, eq,G}(t) \simeq \left[{2.5 \times 10^{13} \over R_0}\right]  ~\dot{M}_{-5}^{1/2} ~V_{\rm w,10}^{1/2} ~
\left({t  \over t_0}\right)^{-m s/ 2} \ ,
\end{equation}
where the stellar mass loss rate is in units  of $10^{-5}
M_{\odot}/\rm{yr}$ and the wind speed is in units of 10 km/s.
For the case of a red supergiant wind where the pre-SN mass loss rate $\dot{M}_{-5} \sim 1$, $V_{\rm w, 10} \sim 1$, and $R_0 \sim 10^{15}$ cm,
field strengths of the order of 25 mG are obtained, far below those shown in the tables.
It requires $V_{\rm w} > 1000$ km/s and high mass loss rates to find $B_{\rm eq} \sim$ Gauss, which may be the case in some WR winds. In general, however, amplification factors of two to three orders of magnitude above $B_{\rm eq}$ are deduced from radio observations.

Recently, \citet{KunduEtal2017} discussed possible constraints
on the circumstellar medium from the radio non-detection of two Type
Ia supernovae, SN 2011fe and SN 2014J. They found a very low-density
medium around both the SNe assuming that about 20\% of the shock
bulk energy was shared equally between electrons and magnetic
fields. Note that the non-linear DSA modeling by
\citet{Bykov3inst2014} predicted about 10\% efficiency of
non-adiabatic magnetic field amplification by CR-driven
instabilities (see Fig.~\ref{fig:PvsU} below), while the electron
acceleration efficiency is usually lower.

As already mentioned,  the physical processes producing magnetic
field amplification are not fully understood.
The possibility that CR driven plasma instabilities  can produce
this amplification is discussed below and we refer to
\cite{Tatischeff09} and \cite{Marcowith14} for additional work.
Before providing more details on the CR instabilities we now briefly
review the  observational evidence that such instabilities develop
in more evolved SNRs.

\subsection{Supernova remnants}\label{S:SNR}
Supernova remnants are known sites of particle acceleration as they produce non-thermal radiation from radio to \gamrays\
\citep[see][for a review]{MarcowithEtal2016}.
Associated with this particle acceleration is clear evidence of
\MFA\  in all historical SNRs making it likely that the two processes are tightly linked \citep[e.g.,][]{Parizot2006}. 

The most direct evidence for MFA comes from high angular resolution observations of nonthermal X-rays.
The {\sl Chandra} X-ray satellite has detected in historical SNRs, in the hard $4-6$\,keV band,
thin filaments of typical size 1-10\% of the remnant radius.
Such X-ray emission can hardly be interpreted as Bremsstrahlung radiation from relativistic electrons as it would required a very high ambient ISM density \citep{Ballet2006}.
Hence, this radiation is interpreted as synchrotron emission produced by multi-TeV electrons.
The size of the filaments puts some constraints on the post-shock magnetic field intensity: electrons downstream from the shock are transported either diffusively or are advected with the flow but can radiate X-rays only within a synchrotron loss timescale.
Since both the diffusion coefficient and the synchrotron loss time
depend on the magnetic field intensity the radiation zone is
limited. Typical field strengths between 50 (in older SNR like SN
1006) and 500 $\mu$G (in younger SNRs like Cassiopeia A) are deduced
by this technique \citep{Parizot2006}. In fact, since  {\sl Chandra}
has a finite point spread function, the deduced magnetic field
intensity is only a lower limit in some cases.

 Lately, {\sl Chandra} has detected regularly spaced stripes in the
X-ray \syn\ emission from Tycho's SNR \citep{EriksenEtal2011}. This
pattern and spacing has been interpreted by \citet{beopu2011} as a
signature of magnetic turbulence produced by the short-scale Bell
instability driven by PeV CR protons. However, recent observations
by VERITAS and Fermi \citep[i.e.,][]{Archambault2017} suggest that
the proton spectrum may be cutting off well below PeV energies. If
the stripes are, in fact, the result of CR-driven turbulence, the
turbulence may result from the long-wavelength mirror instability
driven by CRs undergoing super-diffusion (i.e., \levy\ flight; see
\S~\ref{sec:trans}) \citep[][]{Bykov_SuperD2017}. The turbulence
resulting from the mirror instability will have characteristic
wavelengths long compared to the CR gyroradius, in contrast to the
shorter wavelength turbulence produced by the Bell instability.

An important observational perspective would be to identify filaments at other wavelengths to constrain the MFA process.
The improved sensitivity at 1~GHz of the {\sl Square Kilometer Array} (SKA) (see \S~\ref{S:SKA}), with less than an arcsecond angular resolution, could bring some strong constraints on radio filaments and radio precursors \citep[e.g.,][]{Achterberg94}.  

Supernova remnants are also strong \gamray\ emitters.
This \gamray\ emission  can be of leptonic origin, i.e., \IC\ (IC) or \brems\ radiation in dense environments, or hadronic, i.e., \pion.
A trend seems to appear as a function of the remnant age \citep{Acero15, Acero16}. Young (with an age $< 1000$ yr) SNRs usually show hard spectra with a photon index close to 2 whereas older SNRs exhibit soft spectra beyond GeV energies. The \gamray\ spectra of young SNRs can be interpreted within the framework of one-zone models as being produced by IC radiation. But from X-ray observations we know that in SNRs where MFA
is occurring (see above), a one zone description of acceleration and cooling is likely a poor approximation since the magnetic field will vary behind the forward shock \citep{Marcowith10}.
Also, if SNRs propagate  into the perturbed
ISM or CSM, density inhomogeneities can boost \gamray\ emission
from \pion\ forcing multi-zone models \citep{Gabici14}.

In the few young core-collapse SNRs which are \gamray\ emitters
(e.g., Cassiopeia A,  RX J0852.0-4622), the \gamray\ spectrum seems to cut off around 10 TeV implying a CR spectrum cutting off below $\sim 100$\,TeV.
For the moment there are no hints of a source of PeV CRs which would then produce \pion\ \gamrays\ of $\sim 300$ TeV.
Older SNRs show a cut off below one TeV.
One possibility is that the Pevatron phase is very short after the SN outburst and ends before an age of $\sim 1000$\,yr. This question is treated in the next section.

\section{CR acceleration beyond PeV}\label{S:PEV}
Observations of the spectrum, composition, and anisotropy of CRs
below the knee were briefly mentioned in \S\ref{sec:intro}
\citep[see also][]{Hillas2005}. An important question that is yet to
have a definitive answer for higher energy CRs is where the
transition between the galactic and extragalactic CR components occurs.
It is even possible that the spectral  breaks that have been
observed below PeV energies, combined with the light-mass CR
component between $10^{17} - 10^{18}$\,eV suggested by {\sl LOFAR}
observations, may indicate the presence of different types of
galactic sources which have not been previously considered.

While providing a sufficient energy budget is the most important
constraint for sources of GeV-TeV CRs, the energy content in CRs
beyond the knee is considerably less. This opens the possibility
these CRs may be produced by low-power sources or by rare powerful
events. Of course SNe are still among the possible sources of
high-energy galactic CRs and \citet{Sveshnikova2003} discussed the
role the diversity in supernova explosion energies might have on CR
spectral features around the knee.
In particular, PeV CRs may be produced by a
subset of type Ibc SNe  with \transrel\ shocks
\citep[e.g.,][]{Budnik2008,Chakraborti2011,EWB2013},
or by superluminous SNe
\citep[e.g.,][]{MuraseEtal2011,Murase2014,ZP2016},  or by SNe  in
young compact stellar clusters
\citep[][]{Bykov2014,Bykov_Clusters2017}. We shall now briefly
discuss these sources.

\subsection{CR acceleration in trans-relativistic
supernovae} \label{sec:trans}
\begin{figure}   
\includegraphics[width=5.0in]{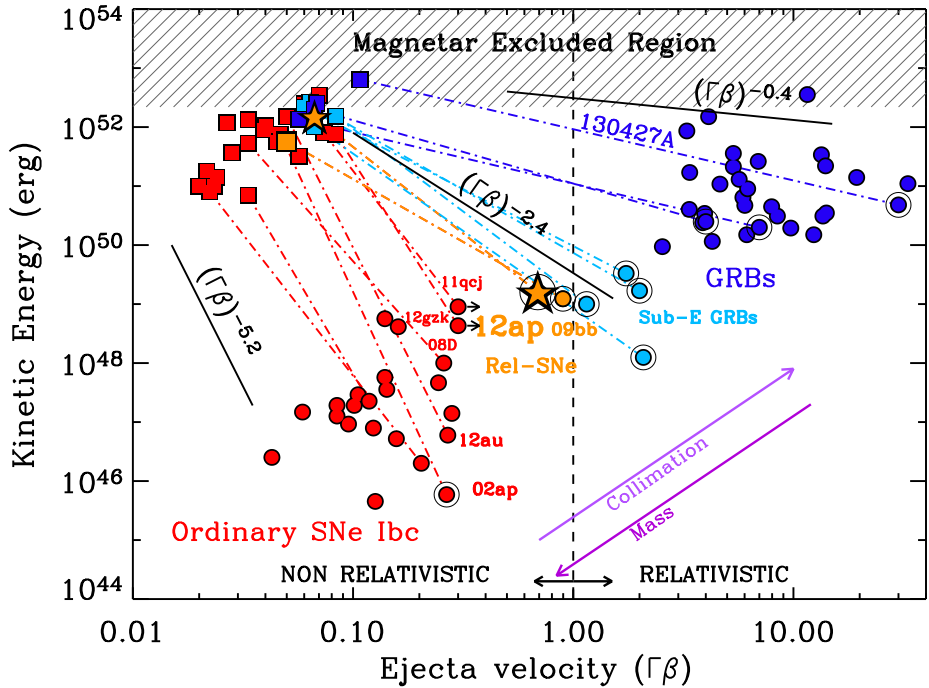}
\caption{The kinetic energy of the ejecta as a function of the
spatial component of the ejecta four velocity $\Gamma\beta$ is shown for distinct
classes of core collapsed events: ordinary type Ibc SNe (red), GRBs
(blue), sub-energetic GRBs (light-blue), and relativistic supernovae
SN 2009bb and SN 2012ap (orange). Adapted from \citet{Margutti2014}.}
\label{fig:Margutti} \vspace{-1.\baselineskip}
\end{figure}

Soon after the cosmological nature of \gamray\ bursts was
established an unusually radio bright supernova SN1998bw was
associated with GRB980425. This SN showed clear evidence for a
blast wave moving at \rel\ speeds  \citep[e.g.,][]{Kulkarni1998}.
The mass in the relativistic ejecta of SN1998bw was estimated as
$M_{\rm ej} \sim 10^{-5} M_{\odot}$, with the energy in the
relativistic outflow $\gsim 10^{49}$ erg. There are also indications
for the existence of \rel\ SNe without a detected GRB. The rare
subclass of type Ibc SNe is distinguished from ordinary SNe by the
presence of ejecta moving at mildly \rel\ speeds
\citep[e.g.,][]{Soderberg2006}.
In Fig.~\ref{fig:Margutti} \citep[adapted from][]{Margutti2014}
the kinetic energy of the ejecta
as a function of the spatial component of the ejecta four velocity
is shown for distinct classes of core collapsed events:  ordinary
type Ibc SNe (red), GRBs (blue),
sub-E GRBs (i.e., sub-energetic; light-blue), and the
relativistic supernovae SN 2009bb and SN 2012ap (orange).
The events lying near the $\Gamma \beta=1$ line represent a subset of SNe with mildly relativistic ejecta speeds and other properties intermediate between ordinary SNe and GRBs.

A prototype of a \rel\ SNe without a detected GRB is SN 2009bb
\citep{Soderberg2010,Margutti2014}. In this remnant, radio
observations were used to probe the fastest ejecta as the expanding
\rel\ blast wave propagated through  the fast wind of the progenitor
star.
In the case of SN 2009bb, \citet{Soderberg2010} estimated that the
ejecta had a minimum energy of $(1.3 \pm 0.1)\times 10^{49}$\,erg,
with a \Lor\ factor $\Gamma \simeq 1.3$ (see Fig.~\ref{fig:Margutti}), and were propagating in the
wind of the progenitor star with an estimated mass loss rate of
$\dot{\rm{M}}= (2 \pm 0.2)\times 10^{-6}~
M_{\odot}~{\mathrm{yr}}^{-1}$.

A simple estimate of the deceleration length, $R_s$, of a SN shock
assuming the ejected mass, $\Mej$, equals the amount of swept up
mass in a progenitor wind with speed $v_w$ and mass loss rate
$\Mdot$ is
\begin{equation}
R_{\rm s} \approx 10^{17}  \left[\frac{M_{\rm ej}}{10^{-5}
M_{\odot}}\right] \left[\frac{v_{\rm w}}{3\times10^3 \kms}\right]
\left[ \frac{10^{-6} M_{\odot}~\mathrm{yr}^{-1}}{\dot{\rm{M}}}\right]~~{\rm
cm} \ .
\end{equation}
The stellar wind mass-loss rate of
$10^{-6} M_{\odot}~\mathrm{yr}^{-1}$,
and the wind terminal speed of $3000$\,\kmps, are consistent with
predictions
for the He main sequence stars which were presented recently by
\citet{Vink2017}, while the Wolf-Rayet type stars
collected by \citet{Nugis_Lamers2000} demonstrated somewhat
higher mass-loss rates and lower wind velocities.
The values and configurations of the  magnetic fields in the winds
of the potential progenitors are not very certain
\citep[e.g.,][]{WalderEtal2012}, while the analysis of radio
observations of SN2009bb \cite[][]{Chakraborti2011} revealed a
magnetic field of $\gsim 0.1$\,G at the estimated radius of $\gsim
10^{17}$\,cm (see their Figure 1 for more details).

It is important to  note that in a \rel\ SN, both the shock speed and magnetic
field in the wind of the progenitor star are expected to be much
higher than in the ISM.
This will strongly reduce the acceleration
time of particles subject to \Facc\ and may make it possible to
accelerate nuclei well beyond PeV energies in a few months.

\citet{Budnik2008}  estimated the conditions for \Facc\ in
\transrel\ SNe and concluded that the estimated rate and energy
production by these  sources are high enough to allow them to power
CRs with energies up to 10$^{18}$\,eV.
They assumed that the magnetic field is amplified by the shock both in
the upstream and downstream regions to values close to equipartition.
Moreover, \citet{Chakraborti2011}  argued that relativistic SNe
similar to SN 2009bb could
 be the sources of ultra-high-energy-cosmic-rays (UHECRs) with energies beyond the Greisen-Zatsepin-Kuzmin limit
(i.e., $\gsim 10^{20}$\,eV) if they have magnetic fields above $\sim
0.1$\,G at radii beyond $\sim 10^{17}$\,cm.

Of course \rel\ SNe are  scarce. The \citet{Chakraborti2011}
conjecture was made assuming that the type Ibc SN rate is $\sim
1.7\xx{4}\,\mathrm{Gpc}^{-3}\,\mathrm{yr}^{-1}$ and the fraction of
these which are mildly \rel\ is $\sim 0.7\%$ (the uncertainty is
large, however, $0.7^{+1.6}_{-0.6}\%$ , or (0.1 - 2.3)\%). If the
fraction 0.7\% is right then one would expect a few \transrel\ SNe
occurring within a distance of a few hundred Mpc every year, and one
every  $5\xx{4}$\,yr in the Milky Way.
A crucial point in these  CR energy estimates is the magnetic field
strength. Strong MFA is simply assumed but this must be justified by
\SC\ models for MFA in \NL\ \Facc\ in \transrel\ shocks.

Relativistic SNe may provide  a distinct component of CRs beyond PeV
energies.
Supernova statistics in  a volume-limited sample based on nearby
SN rates from the Lick Observatory Supernova Search
\citep{LiEtal2011} found that the most abundant type of SNe are
type II (57\% of the total), while type Ibc comprised about  19\%, a
fraction similar to that of SNe Ia which is estimated as 24\%.

If indeed relativistic SNe are just 0.7\% of all type Ibc SNe,  as
suggested by \citet{Soderberg2010} and \citet{Chakraborti2011}, than
their rate in the Galaxy is about one every  $5\xx{4}$\,yr.
The diffusion coefficient $D$ required for the high energy CRs to be
confined in the Galactic halo of size $R$ for 5$\times 10^4$\,yr can
be estimated to be $D \sim 2\times 10^{31}
(R/3\,\mathrm{kpc})^2\diff$.
Note that the gyroradius of a proton of energy $10^{18}$\,eV in the
typical interstellar magnetic field is $\sim 10^{21}$\,cm providing
a minimum  value for  $D \sim 10^{31}\diff$. The minimum value was
derived assuming Bohm diffusion where the CR mean-free-path is equal
to the gyroradius, however, the diffusion coefficient in the halo is
expected to be well above the minimal value.
This value, and the 2\% upper limit on the anisotropy of $\sim
10^{18}$\,eV CRs \citep[see][for a discussion]{Ptuskin2012},
requires  an extended CR halo of size $\sim 10$\,kpc if the
relativistic SNe are to be the dominant source of CRs up to $\sim
10^{18}$\,eV. The CR-driven galactic winds  \citep[e.g.,][and
references therein]{FRYZ17,RBM17} may affect the scale size $R$ of
the CR confinement region.
The power required to maintain this high-energy galactic CR
component (assuming it is quasi-steady) is about 10$^{36}$ \ergs\
meaning \transrel\ shocks would need to place $\sim 10\%$ of the
mildly relativistic outflow kinetic energy into CRs extending to
$10^{18}$\,eV. It has also been suggested
by \citet{LW06} that \rel\ SNe may produce high-energy
CRs in starburst galaxies and be a source of high-energy neutrinos.

The UHECRs with energies well above $10^{18}$\,eV are most probably
of extragalactic origin.  The main source candidates are \gamray\
bursts and powerful radio-galaxies. Galactic magnetars may also
contribute. In all cases, high power and \rel\ outflows are required
to produce the most extreme CR energies
\citep[e.g.,][]{Waxman1995,blasi_magnetar_2000,
Arons2003,Lemoine2013,Asano2016}.

To address MFA with high CR acceleration efficiency in \transrel\
shocks we generalize a Monte Carlo technique that has been used
previously to study \NL\ \Facc\ in shocks of all \Lor\ factors but
without a \SC\ calculation of MFA in \rel\ flows
\citep[][]{EWB2013,WarrenEllison2015,WEBN2017}.
This previous model
assumed a given structure for the magnetic turbulence which was not
directly determined by the CR distribution.
While neglecting MFA in \rel\ shocks might  be acceptable if \Facc\
is inefficient, the efficiency estimate given above suggests that
\Facc\
is efficient enough in \transrel\ shocks to produce  CR pressure
gradients and CR currents strong enough to drive  instabilities
responsible for MFA in a way similar to what is believed to occur in
\nonrel\ shocks
\citep[e.g.,][]{bell_lucek01,Bell2004,BellEtal2013,SchureEtal2012,
bbmo13,MarcowithEtal2016,Bykov_SuperD2017,Marle_Casse_Marcowith2017}.

\newlistDE

The \nonrel\ \mc\ simulation self-consistently finds
a steady-state, planar-shock solution for the \NL\ shock structure including
\listDE injection of particles from the thermal background,
\listDE modification of the precursor flow by energetic particle and magnetic field pressure,
\listDE calculation of the magnetic turbulence (i.e., MFA) due to the anisotropic distribution of the accelerated
particles,
\listDE turbulence cascade,
\listDE a determination of the momentum and space dependent scattering
\mfp\ from the self-generated turbulence,
\listDE a \SC\ calculation of the scattering center speed
\footnote{See \citet{Fiorito1990} for an early calculation of the
scattering center speed using \mc\ methods.} (which may model an
effect of a mean electric field in the shock upstream)
without reference to the \alf\ velocity, and
\listDE particle escape at an upstream \FEB.
Most recently \citep[i.e.,][]{Bykov3inst2014,Bykov_SuperD2017},
super-diffusion (also called \levy-walk or \levy-flight propagation)
has been included within the full \NL\ code structure.

\begin{figure}   
\includegraphics[width=4.0in]{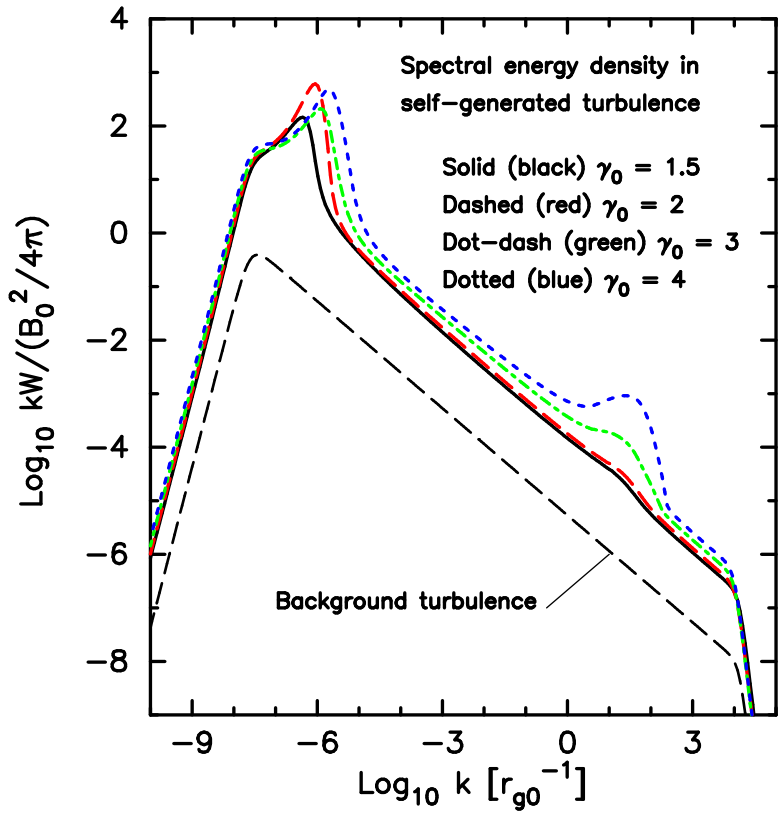}
\caption{Preliminary results  showing
spectral energy densities of the CR-driven magnetic
fluctuations measured in the downstream
region of \transrel\ shocks with indicated \Lor\ factors, $\gamZ$.
The black dashed curve is the background turbulence. There is no
turbulence cascade in this model
\citep[][]{BykovTransRelMFP}.}  \label{fig:TR_MFA}
\vspace{-1.\baselineskip}
\end{figure}

\begin{figure} 
\includegraphics[width=4.5in]{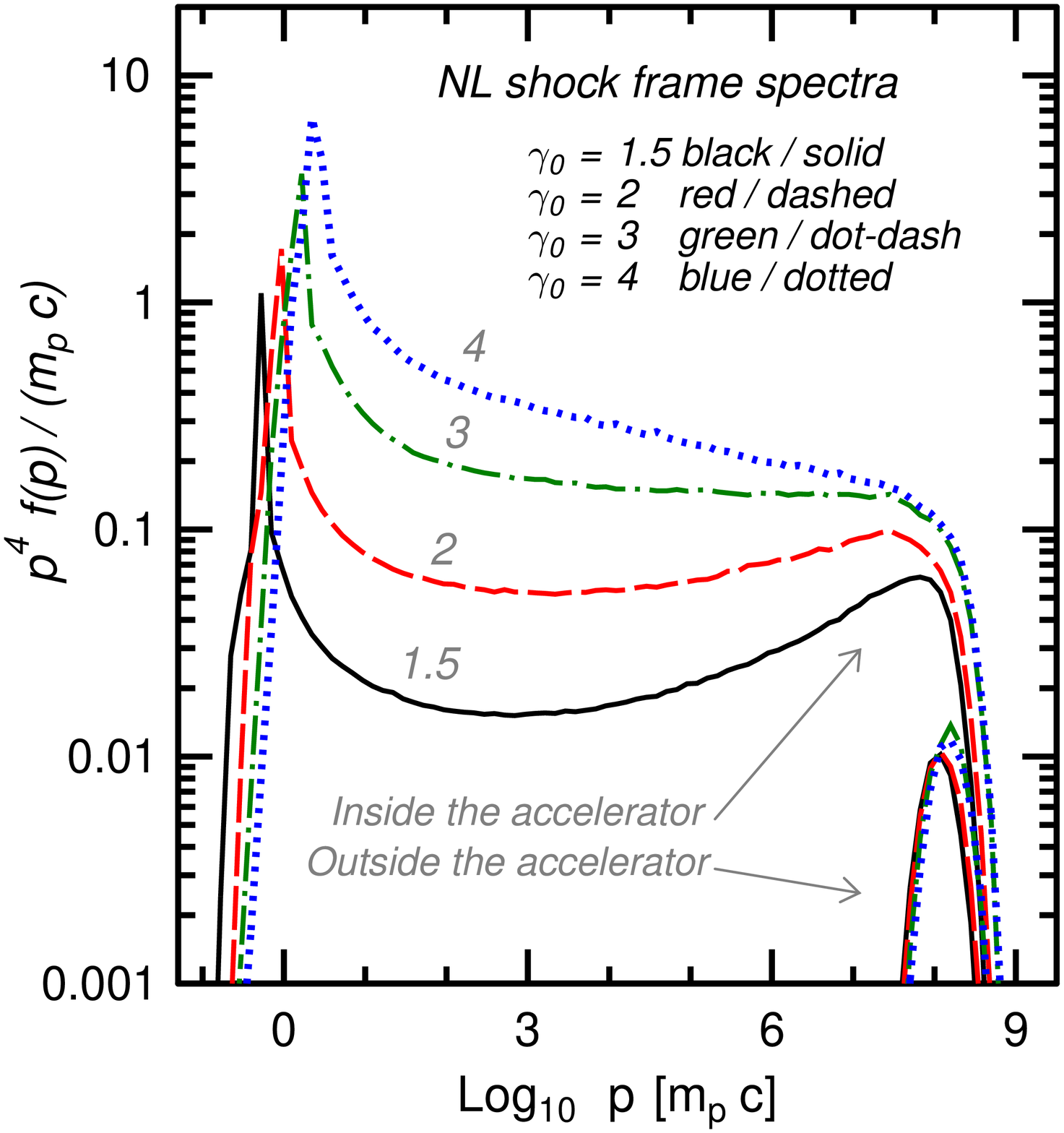}
\caption{Shown are proton phase-space distributions measured  in the
shock rest frame \citep[preliminary results
from][]{BykovTransRelMFP}. These \NL, steady-state, \mc\ simulations
were performed for the \transrel\ shock \Lor\ factors, $\gamZ$,
indicated. The amplified magnetic turbulence determined \SCly\ with
these spectra  is shown in Fig.~\ref{fig:TR_MFA}.} \label{fig:TR_CR}
\vspace{-1.\baselineskip}
\end{figure}

In contrast to PIC simulations, where the magnetic turbulence is
determined directly from particle motion using Maxwell's  equations,
the \mc\ code calculates the resonant and non-resonant instabilities
by coupling analytic descriptions of the growth rates with the
anisotropic particle transport simulated by the code.
In Fig.~\ref{fig:TR_MFA} we show preliminary results from our \transrel\ code.

The figure shows spectra of magnetic fluctuations  for \Lor\ factors
between 1.5 and 4 assuming no turbulence cascade along the initial
magnetic field \citep[c.f.,][]{Lithwick2001}.
The turbulence was calculated \SCly\ with the CR distributions shown
in Fig.~\ref{fig:TR_CR}. It is critical to note that, for clarity,
Figs.~\ref{fig:TR_MFA} and \ref{fig:TR_CR} only show spectra for the
downstream region (the particle distributions are calculated in the
shock-rest frame). There is no strict one-to-one correspondence
between the  CRs and the turbulence in these plots. All spectra are
calculated throughout the shock precursor with the convection of
turbulence and CRs taken into account. The turbulence shown in
Fig.~\ref{fig:TR_MFA} is the net result, in the downstream region,
of production, convection, and cascading, if present.

The effect of strong magnetic field amplification is clearly seen in
Fig.~\ref{fig:TR_MFA}. The self-generated turbulence stands more
than an order of magnitude above the background level (dashed black
curve) for all wavenumbers and \Lor\ factors. In addition, there is
much stronger amplification of long-wavelength turbulence (i.e., $k
\sim 10^{-6}\,\rgz^{-1}$) produced by CRs that escape at the
upstream FEB.
The quasi-thermal CRs also produce enhanced turbulence  for $k \sim
10-100\,\rgz^{-1}$.\footnote{Particle-particle interactions, i.e.,
Coulomb collisions, are not explicitly modeled in the \mc\
simulation so true thermal distributions are not produced.
Therefore, we use the term ``quasi-thermal" to describe the peaks at
thermal energies seen in Fig.~\ref{fig:TR_CR}. These peaks have
essentially the same mean and total  energy as the equivalent \MB\
distributions.}
While not shown, we have performed these calculations with
Kolmogorov-type cascade and found little difference in the particle
spectra or the  total value of the amplified magnetic field.
The background turbulent magnetic field in the wind of the
progenitor star (dashed line in Fig.~\ref{fig:TR_MFA}), integrated
over the spectrum, had a magnitude $\Bturb \sim 0.01$ G for these
simulations. This is considerably less than the shock ram pressure
and the amplified turbulent magnetic field in the shock downstream
which is $\sim 0.2$ G. The wind of the WR type progenitor star with
the magnetic field value $\sim 0.01$ G at R $\sim 10^{17}$ cm is
magnetized if the mass-loss rate is $\sim 10^{-6}
M_{\odot}~\mathrm{yr}^{-1}$.

\begin{figure}  
\includegraphics[width=4.0in]{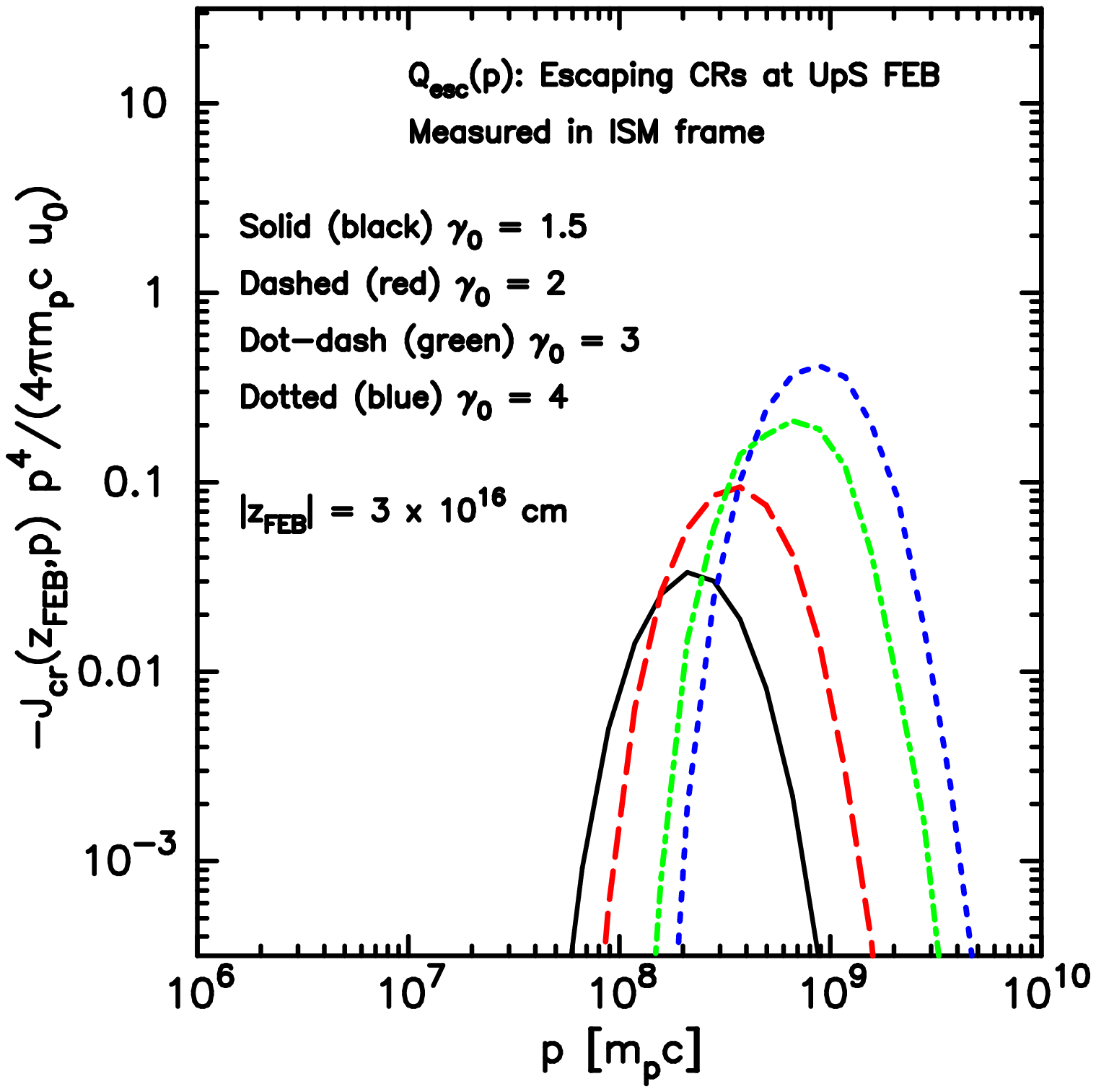}
\caption{Shown are distributions of protons  escaping at the
upstream free escape boundary (FEB)
(i.e., Eq.~\ref{Qesc}), as measured in the far
upstream (i.e., the ISM) rest
frame. The \transrel\ shock \Lor\ factors are indicated.
These are  preliminary results from \citet{BykovTransRelMFP}. }
\label{fig:TR_ESC}
\vspace{-1.\baselineskip}
\end{figure}

We find, even for $\gamZ = 4$, that escaping CRs are important.
In Fig.~\ref{fig:TR_ESC} we show distributions of escaping CRs as seen by a far upstream
observer.
These spectra are determined by
\begin{equation}
\Qesc(p) = - \frac{\Jcr(\zFEB,p)p^4}{4\pi m_p c u_0}, \label{Qesc}
\end{equation}
where $u_0$ is the shock speed and $\Jcr(\zFEB,p)$ is the CR current, both measured in the
far upstream rest frame. Here $\zFEB$  is the position of the upstream FEB and for the plots shown, $\zFEB$ was $3\xx{16}$\,cm from the subshock at $z=0$.

\subsection{CR acceleration in interaction powered luminous supernovae}\label{SLSNe}
Superluminous SNe with a peak luminosity in the optical-UV of
$\sim 10^{44}$\,\ergs\ were recently recognized to be a broad class of
events with the rate estimated to be $\sim 10^{-4}$ of the
core-collapse SN rate \citep[e.g.,][]{BranchWheeler17}.
The detected events were both hydrogen rich and hydrogen poor, while
some of them were apparently dominated by radioactive $^{56}$Ni
decay \citep[e.g.,][]{GalYam2012}.  A superluminous SN event may
occur in the case of a dense enough circumstellar medium around the
SN which provides a radiation-dominated shock propagating through
the envelope
\citep[e.g.,][]{FalkArnett1977,Chugai1994,ChevalierI2011,Chevalier2012,Blinnikov2016,
Blinnikov2016B,SorokinaEtal2016}.
Pre-supernova activity of a massive progenitor star during some
evolutionary phases may have a very large mass-loss rate in the form of
a dense slow wind that  produces dense circumstellar envelopes
\citep{Smith2017,Morozova2017}. Multiple envelopes may be formed if
outbursts occur at different progenitor star evolutionary stages prior
to the SN event.

Assuming that $\sim 10\%$ of the shock ram pressure can be converted
to the fluctuating magnetic field (c.f. Fig.~\ref{fig:PvsU}) and
using the minimal Bohm diffusion coefficient of CRs in the amplified
field, \citet{ZP2016} estimated the maximal energy of
protons accelerated by the Fermi mechanism in type IIn SNe to be
\begin{equation}
\frac{E_{\rm max}}{80~{\rm PeV}}  \leq
\frac{E_{\rm SN}}{10^{52}~\mathrm{erg}}
\left(\frac{M_{\rm ej}}{10~M_{\odot}}\right)^{-1}
\left(\frac{v_{\rm w}}{100 \kms}\right)^{-0.5}
\left(\frac{\dot{\rm{M}}}{10^{-2}~ M_{\odot}~\mathrm{yr}^{-1}}\right)^{0.5}.
\end{equation}
This implies that hypernovae with ejected kinetic energies
of $\Esn \sim 10^{52}$\,erg can accelerate protons beyond PeV energies.
To reproduce the observed light curves of the
superluminous SN 2006gy, assuming that the supernova is powered by
the collision of supernova ejecta with a dense circumstellar medium,
\citet{MoriyaEtal2013} derived an ejecta mass $\lsim 15
M_{\odot}$  and an explosion energy $\gsim 4\xx{51}$\,erg.
\citet{Moriya2014}, based on the observed rise times and
peak luminosities in type IIn SNe, found substantial
diversity in their progenitor wind  densities, SN ejecta
energies, and ejecta masses.

The interactions of the accelerated CRs with the dense circumstellar envelope
of the interaction powered SNe would result in rich
multi-wavelength nonthermal
emission, as well as high-energy neutrinos in the energy range
of the {\sl IceCube Observatory}
\citep[e.g.,][]{Katz2011,Murase2014,ZP2016,PetropoulouEtal2017}.
The pp collisions
between the accelerated CRs and the dense circumstellar envelope
would efficiently produce \gamrays\  and neutrinos.
Moreover, \citet{Murase2014} argued that
secondary electrons and positrons, which are copiously
produced in the inelastic pp collisions, radiate efficiently and would
produce high-frequency synchrotron radio emission which could  be observed by the {\sl Jansky Very Large Array}
and the {\sl Atacama Large Millimeter/submillimetre Array}
from type IIn  SNe at Gpc distances.
\citet{PetropoulouEtal2017} estimated the diffuse neutrino emission from the SN IIn
 to be $\sim 10\%$ of the  neutrino flux above 60 TeV  observed
by the {\sl IceCube Observatory} \citep{Aartsen14}.
They also concluded that to produce
the observed neutrino flux, the high-energy neutrino sources associated
with type IIn events should comprise $\sim 4\%$ of all core collapse SNe
and have a high-energy proton acceleration efficiency $> 20\%$.

\section{Hydrodynamic models of evolving SNRs}\label{s:hydro}
An effective way to model the dynamic nature of SNRs is to couple
the hydrodynamic evolution of the remnant with the CR production. A
number of different models have been presented along these lines and
we refer the reader to the original papers for details
\citep[e.g.,][]{BEK96,EPSBG2007,KJ2009,TDP2011,FDS2012,LEN2012,BTP2016}.
Supernovae and \FoFSA\ have been actively studied for several decades  and any reasonably consistent model of an evolving SNR producing CRs with an efficiency above 5 or 10\% will be complex with a number of parameters and assumptions.
In simplest terms, the different models combine various physical processes
in a more or less \SC\ fashion.
The important physical processes and model assumptions include:

\subsection{Remnant geometry} \label{sec:geo}
Most models assume the SNR is  spherically symmetric but
multi-dimensional models \citep[e.g.,][]{FDS2012} have been
presented. Spherically symmetric models can include dense shells of
material and mimic in an approximate fashion dense clumps of
circumstellar material \citep[e.g.,][]{EB2011}. Multi-dimensional
models are essential for modeling the irregular nature of many
remnants and particularly those, such as SN1006, with clear
asymmetries in emission \citep[e.g.,][]{Cassam2008}.
Besides resulting from irregularities in the CSM, asymmetries, such as that seen in SN1006, may result from a varying shock geometry, as determined by the ambient magnetic field direction. Most theories of \Facc\ suggest that quasi-parallel shocks, those where the ambient magnetic field direction is nearly parallel to the shock normal, inject and accelerate CRs more efficiently than quasi-perpendicular shocks
\citep[e.g.,][]{EBJ95}. Observations and modeling
of SN1006 \citep[e.g.,][]{RothenflugEtal2004,Cassam2008}
support this view.

\subsection{Ejecta profile} \label{sec:ejecta}
The density of material ejected
from the SN explosion is typically assumed to have a power law
\citep[e.g.,][]{ch82} or
exponential spatial distribution in radius \citep[e.g.,][]{DC98}.
A power law distribution, as assumed in Eq.~(\ref{Eq:SSM}), is amenable to \Ss\ solutions
\citep[e.g.,][]{DEB2000} while an exponential distribution may be more appropriate for type Ia thermonuclear SNe. Of course, the ejecta distributions from actual SNe are likely to be more complex, as discussed in the next section.

\begin{figure} 
\includegraphics[width=5.0in]{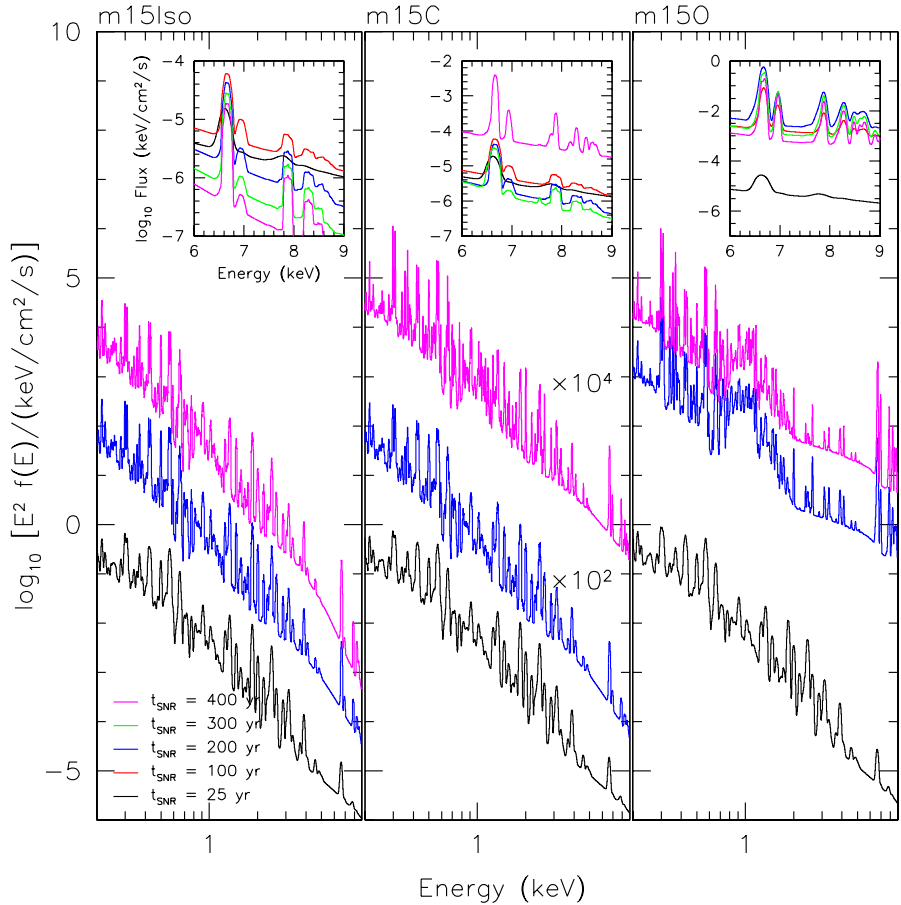}
\caption{Integrated X-ray spectra for three
different end-to-end (i.e., pre-main sequence, through explosion, into the remnant phase)
core-collapse SN models. In all panels X-ray spectra are plotted for different ages, as indicated by color. In the main panels, the spectra at 200 and 400 years are scaled relative to the spectra at 25 years. The inserts show absolutely normalized emission around the Fe-K line. This work shows that X-ray emission, even 100's of years after the explosion, differs noticeably depending on the progenitor mass-loss history and other details of the explosion. This is Fig.~8
from \citet{PatnaudeEtal2017} and this paper should be seen for details.
\label{fig:Xray}}
\vspace{-1.\baselineskip}
\end{figure}

\subsection{Circumstellar material} \label{sec:CSM}
The ejecta and outer blast wave  will interact with the \CSM\ adding
another level of complexity \citep[e.g.,][]{Raymond2017}. While type
Ia SNe may explode in a uniform medium, core-collapse SNe are likely
to explode in a more complex environment produced by pre-SN stellar
winds (i.e., wind-blown bubbles) \citep[e.g.,][]{Dwarkadas2005},
nearby dense molecular clouds, and/or colliding plasmas from nearby
SNe and SNRs \citep[e.g.,][]{Bykov_Clusters2017}. If the SN blast
wave interacts with dense external material, strong \gamray\
emission from proton-proton interactions is expected as in  SNRs
W44, IC 443, and 3C 391 \citep[see
e.g.][]{Tavani2010,uchiyamaea10,AbdoEtalW442010,AbdoEtalIC4432010,AckermannSci13,SlaneSSRv15}.
Recent work by \citet{PatnaudeEtal2017}  has included ejecta
profiles obtained from stellar evolution codes of core-collapse SNe
and demonstrated that the mass-loss history leaves an imprint on
X-ray emission lasting days to years after the collapse (see
Fig.~\ref{fig:Xray}).

\subsection{Production of a forward and reverse shock pair}
In young SNRs  the reverse  shock will co-exist with the forward
shock for a period of time and both will heat the thermal plasma to
X-ray emitting temperatures and simultaneously produce CRs. Except
for the extreme remnant limb, the observed line-of-sight will pass
through material with very different parameters and shock histories.
While single component models are often used, care must be taken to
ensure the multi-component nature of the remnant is accurately
accounted for. Interpreting the integrated  multi-component emission
in terms of simple power laws, as is often done, may be misleading.

\subsection{Effect of CR acceleration on remnant dynamics, plasma heating, and thermal X-ray production}
If CRs are produced efficiently  (i.e., $>10\%$ of the shock bulk
flow kinetic energy is placed in \rel\ particles), the backpressure
of CRs will influence the remnant hydrodynamics since \rel\
particles produce less pressure for a given energy density than
\nonrel\ ones.
The energy placed in CRs comes from the thermal plasma so the shocked temperature is less than expected for \TP\ acceleration. Furthermore, the efficient production of CRs can result in an increase in the shock compression ratio from standard \RH\ values.
The change in temperature and density of the shocked plasma will modify the X-ray line emission. This effect has been studied extensively with a code coupling the remnant hydrodynamics with efficient CR production
\citep[see][and references therein]{ESPB2012,PatnaudeEtal2017}. The effect of CR production influences the \NEI\ X-ray emission in observable ways.

\begin{figure} 
\includegraphics[width=4.0in]{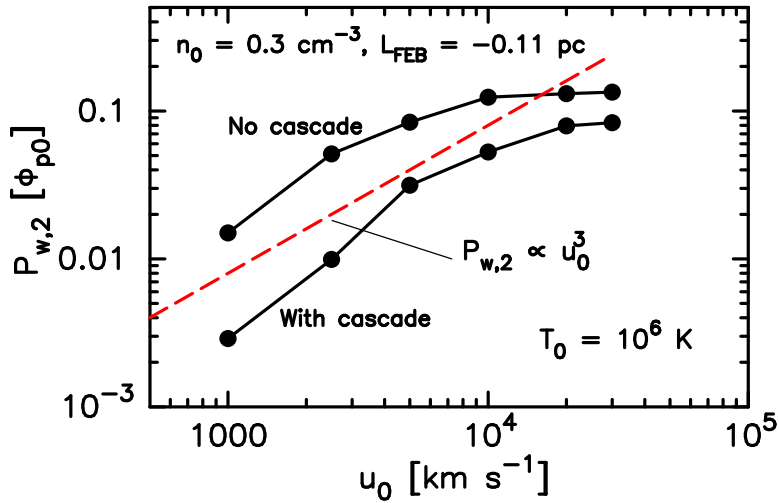}
\caption{The downstream pressure in magnetic fluctuations in units
of the far upstream ram pressure vs. shock speed $u_0$ for various
shock models simulated with a non-linear Monte Carlo model
\citep[see][for details]{Bykov3inst2014}. Models with and without
turbulence cascade are included as indicated. The dashed (red) line
approximates the behavior for low shock speeds. At higher speeds the
turbulence saturates at the $10-15\%$ level. \label{fig:PvsU}}
\vspace{-1.\baselineskip}
\end{figure}

\subsection{Magnetic field amplification (MFA)} \label{sec:MFA}
The self-generation of magnetic  turbulence is necessary for
first-order Fermi shock acceleration to produce CRs to high
energies.
Beyond that, as shown in Section~\ref{sec:radio},
it has become clear from observations of
young SNRs that not only must turbulence be generated, it must be
generated far more efficiently then envisioned when the Fermi
mechanism was first proposed \citep[e.g.,][]{Parizot2006}. This
MFA is strongly \NL\ and is generally added to hydro models
in some approximate way.
Obtaining \SC\ descriptions of MFA from  both resonant and
non-resonant instabilities is an active area of current research, as
discussed in \S~\ref{sec:trans}.

In Fig.~\ref{fig:PvsU} we show figure 11 from
\citet{Bykov3inst2014} giving the pressure in magnetic turbulence
versus shock speed $u_0$,
as obtained with \mc\ techniques including
the resonant CR-streaming instability and two non-resonant
CR-current instabilities
\citep[i.e.,][]{Bell2005,bbmo13}.
The \mc\
results show that the efficiency of MFA, as defined by the pressure
in turbulence, saturates at $\sim 10-15\%$ of the far upstream bulk
flow ram pressure.

\subsection{Cosmic ray escape in \Facc} \label{sec:escape}
For the Fermi mechanism to work,  particles must be confined to the
shock by magnetic turbulence. The scale of this confinement is $\sim
D(p)/u_{\rm sh}$, where $D(p)$ is the diffusion coefficient of a CR
with momentum $p$ and $u_{\rm sh}$ is the
shock velocity.\footnote{Note that the far upstream shock speed is written as either $u_0$ or $\usk$.}
Due to the
self-generated turbulence, this diffusion coefficient is typically
much smaller than that of the quiet ISM
\citep[e.g.,][]{Bell78a,LC83,Bell2004}, i.e., CR driven instabilities can
provide amplification of the  seed interstellar turbulence and, in
some energy ranges, reduce the diffusion coefficient by orders of
magnitude.

However, regardless of how efficient  the MFA is, as long as the shock age is long compared to the acceleration time,  there will always
be some CR energy, determined by shock geometry, above which the
accelerated particles can no longer generate enough turbulence to
confine themselves to the shock.
These high-energy particles will
escape the accelerator while lower-energy CRs remain
confined.\footnote{This assertion depends only on a diffusion
coefficient which is an increasing function of CR momentum and the
fact that all real shocks are finite in extent. In this case, at
some $p$, $D(p)/\usk \sim \Rsk$, where $\Rsk$ is the shock radius,
and the CR can no longer be confined independent of any plasma
physics details \citep[see][for a discussion of effects of the
background ISM field]{Drury2011}.}
If \Facc\ is efficient, the modified compression ratio for the highest energy CRs can become greater than four  \citep[see][]{BE99} and the escaping energy flux can be a significant fraction of the total shock energy flux
\citep[see][for
further discussion]{CBA2009,Drury2011,EB2011,
Kang2013,MalkovEtal2013}.

A consequence of escape is that, if  SN shocks are producing
galactic CRs by \Facc, an outside observer would see at each
particular moment just a relatively narrow spectrum of the escaped
particles centered at some maximum momentum $p_{\rm max}(t)$.
However, since $p_{\rm max}$ evolves with the SNR expansion, the
time integrated CR spectrum can be an extended power law
\citep[e.g.,][]{PZ2005a}.

\begin{figure}  
\includegraphics[width=4.0in]{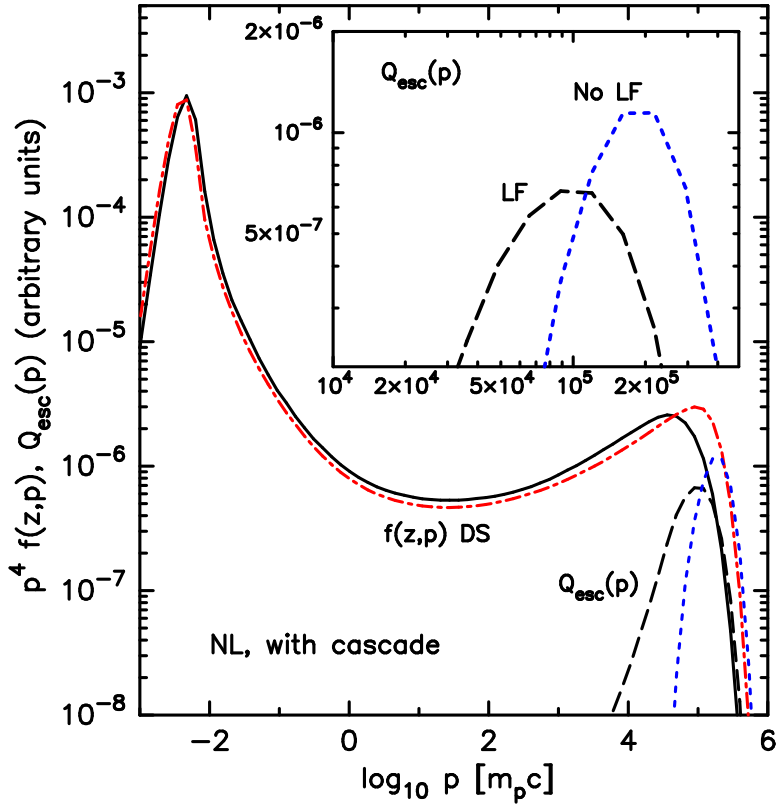}
\caption{Shown are proton phase-space distributions simulated with a nonlinear Monte Carlo model of superdiffusive shock  acceleration
\citep{Bykov_SuperD2017}. Downstream (DS) spectra, as well as the distributions of particles escaping the upstream FEB,
are plotted. In the main panel, the black curves are with superdiffusion and the red and blue curves are with normal diffusion. The insert shows the escaping fluxes with superdiffusion (labeled `LF') and with normal diffusion (labeled `No LF').
All spectra are absolutely normalized relative to each
other.
\label{fig:LF}}\vspace{-1.\baselineskip}
\end{figure}

Basic \Facc\ assumes that fast  particle transport in the shock
precursor is described by standard diffusion due to particle
scattering by magnetic turbulence.
However, the
transport of energetic particles in the presence of
intermittent (i.e., non-Gaussian) magnetic turbulence produced by
anisotropic CR distributions can be different from standard
diffusion \citep[see][for a recent review]{Zimbardo2015}.
This superdiffusive propagation (also called \levy-walk or \levy-flight) modifies the spectrum of escaping CRs. A nonlinear Monte Carlo model of \Facc\  with MFA, including  superdiffusive transport in the
shock precursor, has been presented by \citet{Bykov_SuperD2017}.
In Fig.~\ref{fig:LF} we illustrate the effect of superdiffusive particle transport on the shape of the spectrum of escaping CRs with
superdiffusion (labeled `LF') and without (labeled `No LF').
The escaping CR spectrum is clearly broadened when  superdiffusion is taken into account. This broadening will influence the \pion\
emission and measurements of \gamray\ spectra produced by CRs escaping a SNR adjacent to a molecular cloud can be used to constrain
superdiffusive transport models.

\begin{figure} 
\includegraphics[width=4.0in]{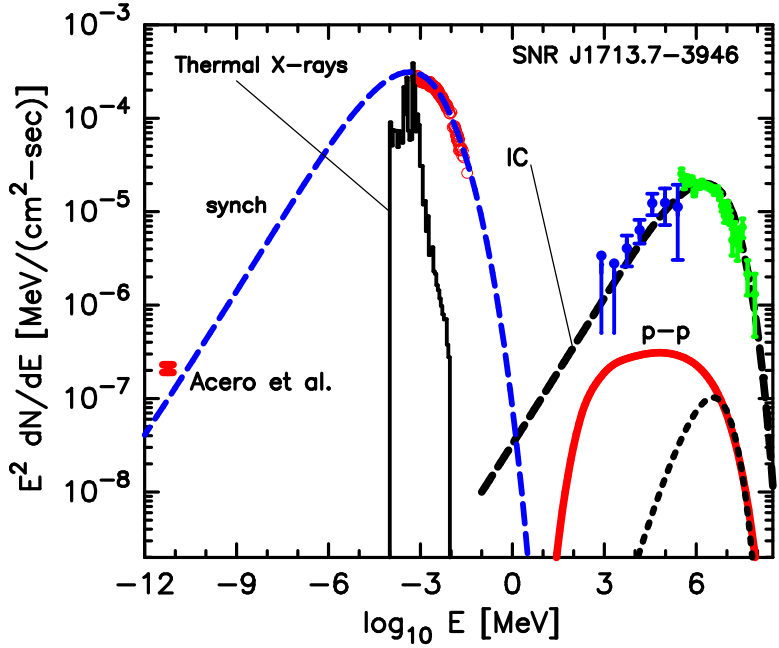}
\caption{Fit to SNR J1713 continuum and X-ray emission line
observations. The different emission processes are indicated except
the dotted black curve which is p-p emission from escaping CRs. This
figure is adapted from figure 2 in \citet{ESPB2012} and that paper
should be seen for details and references to the observations.
\label{fig:J1713}} \vspace{-1.\baselineskip}
\end{figure}

\begin{figure} 
\includegraphics[width=5.0in]{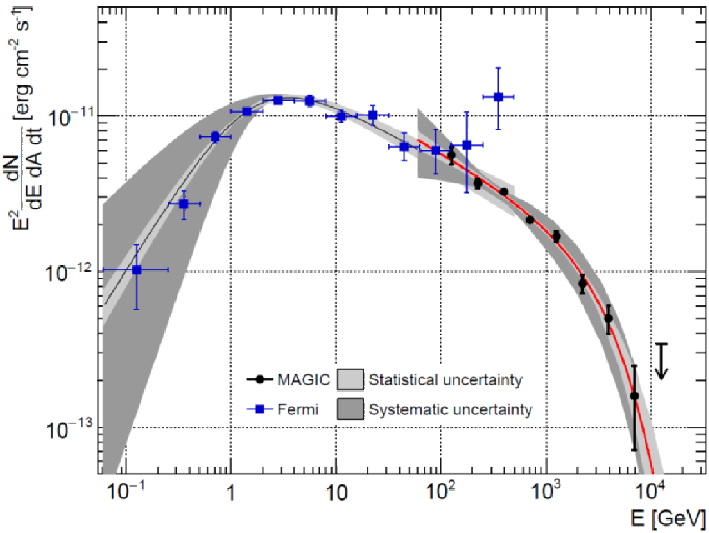}
\caption{Figure~1 from \citet{MAGIC_CasA2017} (``A cut-off in the
TeV gamma-ray spectrum of the SNR Cassiopeia A") showing the
spectral energy distribution from the \CasA\ SNR. As indicated, the
black dots are from the MAGIC telescopes while the blue squares were
measured by the {\sl Fermi} spacecraft. The cutoff at low energies
is an indication of the kinematic threshold for pion production,
while the break at high energies, if this emission is from \pion,
indicates a break in the energy spectrum of hadrons produced by the
SNR blast wave  \citep[see][]{Funk2015}. \label{fig:CasA}}
\vspace{-1.\baselineskip}
\end{figure}

\subsection{Broad-band continuum radiation} \label{sec:broad}
The shock-heated plasma and \rel\  CRs produce radiation that must
be described  if particular SNRs are to be modeled. Since the
thermal X-ray line emission is coupled to the CR production through
\NL\ Fermi acceleration, the broad-band continuum emission must be
determined consistently with the thermal emission.
The continuum processes that must be considered are \synch, \IC\
(IC), and \brem\ from \rel\ electrons, and \pion\ from \rel\ ions
interacting with the background material.
Fitting the broad-band emission with a single set of parameters can
strongly constraint models. This is particularly true for \syn\
because of the large lever arm between radio (from $\sim$ GeV
electrons) and X-ray \syn\ emission (from $10-100$\,TeV electrons).
Care must be taken  of course to ensure the observations from
different instruments with different fields of view, energy
resolutions, etc. are compatible.

In  Fig.~\ref{fig:J1713} we show a fit obtained by \citet{ESPB2012}
to the broadband emission observed from \SNRJ.  The various emission
processes are indicated and for this remnant IC emission dominates
over \pion\ (p-p) at TeV energies. In other SNRs, e.g., Tycho, W44,
and IC~443, \pion\ is seen to dominate. As shown by
\citet{CSEP2012}, there are also cases, such as CTB 109, where IC
and \pion\ contribute almost equally to the \gamray\ emission.

\FFoFSA\ naturally puts more energy into hadrons than leptons. This
is because \Facc\ taps the mechanical energy  of the shock with
near-elastic scatterings between the converging upstream and
downstream flows rather than electromagnetic energy -- heavy
particles get more energy than light ones. All consistent models of
\Facc\ show this.
On the other hand, leptons radiate far more  efficiently than
hadrons. In any particular SNR, environmental factors, most
importantly the density and magnetic field of the \CSM, determine
which of the two competing processes, IC or \pion, dominate.

Supernovae that explode in a dense environment, such as W44 and
IC~443, are likely to show  strong \pion\ emission. If the remnant
is in a low density region, such as \SNRJ, IC is likely to dominate.
While specific features in the photon spectrum in a limited energy
band (such the low-energy turnover shown in Fig.~\ref{fig:CasA} for
Cas~A) can  point to one mechanism over another, broadband fits from
radio to \gamrays, which depend on both leptonic and hadronic
emission, produce stronger constraints. This is particularly true
when thermal X-rays are coupled to CR ion production, as in the fit
shown in Fig.~\ref{fig:J1713}.
It is important to emphasize that, in \nonrel\ shocks at least, even
if IC emission dominates, all consistent theories of \Facc\ show the
underlying shock putting a large majority (e.g., 99\%) of energy
into hadrons.

\section{Supernovae in clusters and clustered supernovae}\label{S:OTH}
Massive stars which are the progenitors of core-collapse SNe are
typically born in dense cores of molecular clouds
\citep[e.g.,][]{Lada2003,Krumholz2017} and therefore are spatially and
temporarily correlated. This correlation  is important for global
models of the interstellar medium \citep[e.g.,][]{heiles90,Cox2005}
and influences many aspects of CR physics.

Observations have revealed massive stars in both dense young compact
clusters, like \West\ with a mass close to 10$^5~\Msun$
\citep[e.g.,][]{Clark2008}, and in  unbound OB associations with
stellar volume densities below 100 stars pc$^{-3}$, such as Cygnus
OB2 \citep[e.g.,][]{WrightEtal2014}.

\subsection{Cosmic rays in superbubbles}
Powerful stellar winds and supernovae in OB associations had been
predicted to be sources of CRs
\citep[e.g.,][]{cm83,BT1990,Bykov2001,Lingenfelter17}. During the
early period of imaging atmospheric Cherenkov telescopes, {\sl
HEGRA} detected an excess of TeV emission with a hard spectrum
spatially coincident with the Cyg OB2 region
\citep{AharonianEtal2002}. The authors discussed the possibility
that this \gamray\ emission originated from TeV particles
accelerated by multiple young massive stars and SNe in the Cyg OB2
region.

More recently, an extended ($50-200$\,pc wide) \gamray\ source was
discovered with the {\sl Fermi Large Area Telescope} in the Cygnus X
region, a giant
 complex of molecular clouds and star-forming regions located at an estimated distance of $\sim 1.4$\,kpc \citep{RyglEtal2012}.
The source (dubbed the Cygnus cocoon)  was identified by
\citet{AckermannSB2011} as a possible superbubble filled with
freshly accelerated CRs.
The hard emission from the Cygnus cocoon extends to 100 GeV with a
flux of $(5.8 \pm 0.9)\times 10^{-8}$ ph cm$^{-2}$ s$^{-1}$ in the
$1-100$\,GeV range. At a distance of 1.4\,kpc, this corresponds to a
\gamray\ luminosity of $(9 \pm 2)\times 10^{34}$\,\ergs.
This \gamray\ luminosity is below 1 percent of the kinetic power of
the stellar winds in Cyg OB2, the rich OB star association located
in the direction of the Cygnus cocoon.

Recently, a second extended {\sl Fermi LAT} \gamray\ source with a
hard spectrum in the GeV range,  possibly associated with the star
forming region  G25.0+0.0, was reported by \citet{Katsuta2017}.
The authors estimate the \gamray\ luminosity of G25.0+0.0 to be
about 10 times larger than that of the Cygnus cocoon for otherwise
similar parameters.
This implies much higher efficiencies of particle acceleration
and/or radiation in the G25.0+0.0 OB association. Given the total
inelastic cross section of proton-proton interaction to be about 30
mb above a few GeV, the proton cooling time would be $\sim
3\xx{7}/n$\,yr, where $n$ is the ambient density in cm$^{-3}$.
If before being released into the surrounded dense shell, GeV-TeV
CRs are confined in a  superbubble filled predominantly with a hot
tenuous plasma, the radiative efficiency of the superbubble will be
low even if the particle acceleration efficiency is high.

The age of an OB association is important for estimating the kinetic
power available from SNe  events. Using a population synthesis
approach \citet{Martinea10} estimated the mechanical luminosity over
the first 3 Myr of the life of Cyg OB2 to be $\sim 4\xx{38}$\,\ergs.
Cygnus OB2 is part of the Cygnus X region and is a rich OB association with many hundreds of OB stars \citep[e.g.,][]{Knodlseder2000,Wright2015}.
\citet{WrightEtal2014} analyzed a  selected sample of {\sl Chandra}
X-ray observations  of young stars and concluded that Cyg OB2 formed
as a highly substructured, unbound  association with a low volume
density ($\leq$ 100 stars pc$^{-3}$).
They found no signs of dynamical evolution of Cyg OB2 which is
inconsistent with the idea that all stars form in dense, compact
clusters.
On the other hand, \citet{ComeronEtal2016} investigated the past
star formation history of Cygnus OB2 using the red supergiants
detectable in the near infrared as a probe to trace massive stars
with initial masses between 7 and 40 $M_{\odot}$. They concluded
that Cygnus OB2 has a history of star formation extending into the
past for at least 20~Myr.
This age is such that star formation started long before the latest
star formation burst which produced the dense aggregate of O-type
stars currently dominating the appearance of Cyg OB2.

Multiple SNe and powerful winds of early-type stars have been
suggested as favorable sites of CR acceleration \citep[see][for a
review]{Bykov2014}.
Extended superbubbles, filled with hot X-ray emitting gas, can be
created by multiple clustered SNe over a time scale of $\gsim 10^7$
years.
The bubbles will contain an ensemble of MHD shocks which may be able
to strongly amplify the  turbulent magnetic fields within the
superbubble leading to a  high efficiency (i.e., $\gsim 10\%$) for
converting kinetic power to freshly re-accelerated CRs.
%
This may result in a substantial temporal evolution of the CR
spectra over times on the order of  10 million years.
Nonlinear modeling of this process \citep[i.e.,][]{Bykov2001}
predicted the time asymptotic CR spectrum to be a power law of index
close to 2 in MeV-TeV range where the energy independent turbulent
diffusion dominates the CR propagation inside the superbubble.
This is consistent with the spectrum of \gamrays\ observed  in the
Cygnus cocoon if  the main radiation mechanism is pion production
from CR hadron  interactions.

The Fermi shock acceleration mechanism assumes that CRs are confined
relatively close to the shock due to a small diffusion coefficient.
This strong scattering is likely provided by the turbulence produced
by CR-driven instabilities, i.e., strong magnetic field
amplification \citep[e.g.,][]{SchureEtal2012}.
This  strong turbulence will also increase the maximum energy CRs
can obtain in the shock. Observations show that the maximum particle
energies in some young isolated SNRs with shock velocities $\gsim$
1,000 km s$^{-1}$ are typically above a TeV.

The bulk of CRs observed at Earth have GeV energies and these must
escape from the SNR at late stages when the CR-generated turbulence
has weakened. For core-collapse SNe, this escape is expected to
occur in a superbubble. These CRs are expected to be confined and
re-accelerated by multiple shocks in the superbubble
\citep{Bykov2001,parizot04,Ferrand2010,Bykov2014}.
Recently, the {\sl Advanced Composition Explorer CRIS} instrument
discovered $^{60}$Fe nuclei in CRs in the energy range
$195-500$\,MeV per nucleon \citep{Binns_60Fe_2016}. The short $\sim
2.6$\,Myr lifetime of $^{60}$Fe suggests these nuclei originated in
nearby clusters of massive stars.

It is worth noting that the observed \gamray\ emission from the
starburst galaxies NGC 253, NGC 1068, NGC 4945 and M82
\citep[see][]{Ohm2012,ohm16} show spectra with photon indexes $\sim
2$. This  is consistent with that observed in both the Cygnus cocoon
and G25.0+0.0 but  much flatter than that of the Milky Way. This may
indicate that superbubble type objects are the dominant CR sources
in starburst galaxies or, alternatively,  that CR transport in
starburst galaxies is energy independent at least up to TeV
energies.
Furthermore, since superbubbles  are adjacent to their parent
molecular complexes, the enhanced density of low-energy CRs they
contain may provide ionization and heating of the dark molecular
clouds. The CR energy deposition may keep the gas temperature
at $\sim 10$\,K and provide the gas ionization fraction $\sim 10^{-7}$
required by current models of the molecular chemistry in dark clouds
\citep[e.g.,][]{Grenier2015}.

High-energy neutrinos produced by inelastic nuclei collisions in the
Cygnus X  region  may have a large enough flux to be detected with
the {\sl IceCube Observatory}, as  estimated by
\citet{Yoast-Hull2017} using a single-zone model of CR  interactions
with the molecular gas.
The planned observations of the Cygnus region with the high
sensitivity and good angular resolution {\sl Cherenkov Telescope
Array} have the potential to provide imaging and spectra of the
region between  a few tens of GeV up to $\sim 100$\,TeV
\citep{WeinsteinEtal2015}. Highly informative MeV to GeV
observations of the region can be performed with
 the planned {\sl e-ASTROGAM} mission \citep[][]{DeAngelis2017}.

\subsection{Cosmic ray acceleration by supernovae in compact clusters}
Compact clusters of young massive stars are sites of SN explosions.
Contrary to loose OB associations,  young stellar clusters have
large core star densities $\gsim 10^{3} M_{\odot} \mathrm{pc}^{-3}$,
with a total cluster mass $> 10^{4} M_{\odot}$ within a virial
radius $\sim$ 1 pc \citep[e.g.,][]{Portegies_Zwart2010}.
The total mechanical power of the OB stars in such a cluster may
exceed $5 \times 10^{38}$\,\ergs\ and will likely launch a powerful
cluster wind,  as modeled by \citet{chev_clegg85} for
starburst-galaxy nuclei.

\begin{figure}  
\includegraphics[width=4.5in]{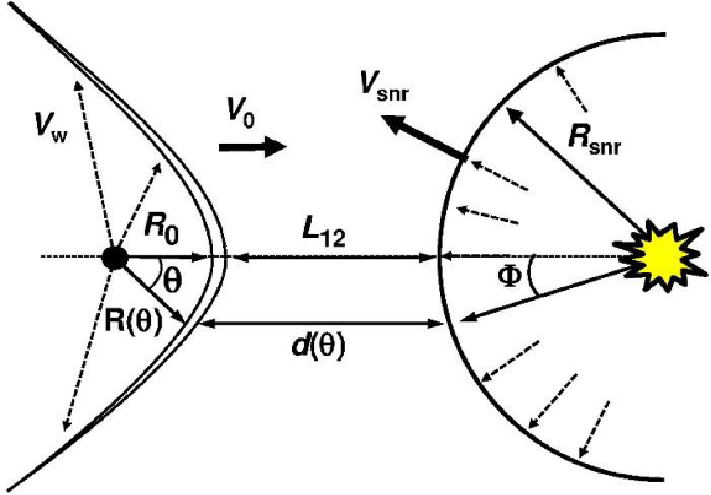} 
\caption{A sketch illustrating the colliding shock flow geometry
where the star cluster wind  is impacted by a supernova shock. See
\citet{BEGO2015MNRAS} for details.} \label{fig:CollSF}
\vspace{-1.\baselineskip}
\end{figure}

\begin{figure}  
\includegraphics[width=4.25in]{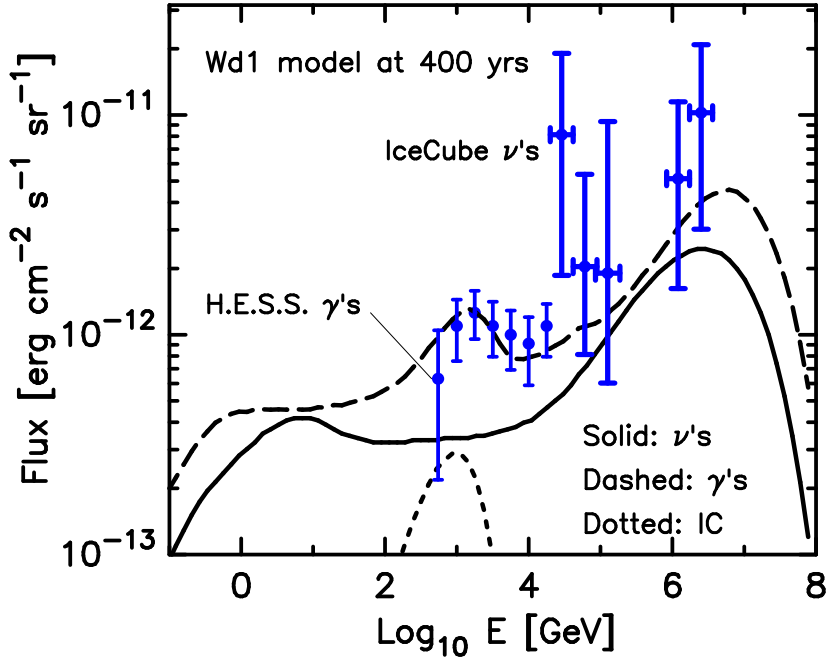}
\caption{Model predictions of \gamrays\  (dashed curve) and
neutrinos (solid curve) from proton-proton interactions calculated
in a colliding shock flow source of age $\sim$ 400 yrs as given by
\citet{BEGO2015MNRAS}. The dotted curve is the IC emission from
primary and secondary electrons accelerated directly in this source.
The extreme upward curvature in the neutrino spectrum above $\sim
10$\,TeV reflects the transition from CR acceleration in the single
SNR shock for low-energy particles to the more efficient
acceleration for high-energy particles as they scatter back and
forth between the SNR shock and the cluster wind. The data points
for the H.E.S.S. source, and the five {\sl Ice Cube} events, are
presented just to illustrate the model predictions.} \label{fig:Wd1}
\vspace{-1.\baselineskip}
\end{figure}

A supernova in such a cluster (see sketch in Fig.~\ref{fig:CollSF})
may be an efficient CR accelerator. As shown by
\citet{Bykov2014} and \citet{BEGO2015MNRAS}, a SN blast wave, colliding with a
fast wind from a compact cluster of young stars, allows the
acceleration of protons to energies well above the standard limits
of \Facc\ in an isolated SN.
The proton spectrum in such a wind-SN PeVatron accelerator is hard
with a large flux in the high-energy-end of the spectrum producing
copious \gamrays\ and neutrinos in inelastic nuclear collisions.
In Fig.~\ref{fig:Wd1} we illustrate the model predictions for the
\gamray\ (dashed line) and neutrino spectra (solid line) produced by
a SN in a cluster similar to the galactic \West\ cluster
\citep[see][for details]{BEGO2015MNRAS}.
It is estimated that the \West\ cluster may accelerate protons
to $\gsim$ 40 PeV and result in enough neutrino production to
account for a few events detected by the {\sl Ice Cube Observatory}
from the inner Milky Way direction.

\section{Prospects for future observational facilities}\label{S:OBS}

\subsection{The Square Kilometer Array}\label{S:SKA}
The {\sl Square Kilometer Array} (SKA, http://skatelescope.org/) is the next high sensitivity radio telescope scheduled for 2020 to be located in South Africa and Australia. The SKA will have an effective surface of 1 km$^2$ with unprecedented sensitivities
of 3.36 and 0.75 $\mu$J/$\sqrt{\mathrm{Hz}}$ for continuum emission at two wavebands $0.05-0.35$ GHz (SKA-Low) and $0.35-14$ GHz (SKA-Mid), respectively. The SKA-Mid will have an angular resolution of 0.25 arcsecond at 1 GHz.
As we have emphasized above, radio observations are essential for understanding many aspects of SN physics and the CSM and
the improved sensitivity of the SKA should permit a breakthrough in SNe radio observations.

\subsubsection{Radio observations of type Ia supernova explosions}
As of yet, no type Ia SN explosion has been observed in the radio band. Hence, the high sensitivity of the SKA may lead to this discovery as well as the ability to follow in time the radio emission of such objects.
The SKA-mid configuration should reach flux sensitivities below the upper limits obtained by the VLA for the most luminous type Ia SNe like SN 2014J or SN 2011fe.\footnote{The most recent anticipated instrumental performance for SKA1 is given in
\texttt{https://astronomers.skatelescope.org/wp-content/uploads/2017/10/SKA-TEL-SKO-0000818-01 }\!
\texttt{\_SKA1\_Science\_Perform.pdf.}}
A SNIa radio detection will help to determine if the stellar system prior to the explosion is double or single degenerate. Radio emission is expected from a single degenerate case as a result of the interaction with the CSM deposited by the companion star \citep{Wang15}.

\subsubsection{Radio observations of core-collapse supernova explosions}
The term core-collapse supernova (CCSN) includes a wide variety
of sub-types, the names of which convey little or no information on the underlying properties of the object. For example, type IIP SNe (about half of the SNe), type IIL, IIb, Ib/c, or IIn \citep[see][for details]{Smith14}.
At present, about 50 CCSN explosions have been detected and followed in time at radio wavebands \citep{Weiler02, Perez15}.
The SKA, with its improved  sensitivity at different wavelengths and
combined pointed and survey modes, is expected to detect several
thousand CCSNe \citep{Perez15} up to a redshift $z\sim 0.25$ for the
most intense type IIn SNe.
With the advent of the SKA  the community will be able to start
statistical analysis of CCSN properties depending of their subclass
type.

The SKA will also be able to  monitor the closest and most intense
events as it was already done with the VLA for emblematic objects
like SN1987A and SN1993J. Time evolution of the radio spectrum is of
particular importance to understand shock dynamics, magnetic field
generation, particle acceleration in different CSM environments.
These issues are intimately related to the problem of the origin of
cosmic rays and make these observations particularly relevant to
trigger multi-wavelength observation campaigns involving high-energy
instruments like the X-ray satellites XMM-Newton or Chandra and
\gamray\ telescopes like the future Cherenkov Telescope Array (see
next sections).

\subsection{The Large Synoptic Survey Telescope}
The Large Synoptic Survey Telescope  (LSST) (https://www.lsst.org/)
is a wide-field survey telescope with an 8.4-meter primary mirror
currently under construction in Chile with a first light expected in
2019. It is sensitive in the 320-1050 nm band and will scan the
entire visible (mostly southern) sky every few nights.
The LSST is an important telescope  for SN optical monitoring and
should obtain photosphere emission lightcurves and act as a trigger
for other multi-wave length observatories. The LSST survey is
designed to detect more than 100,000 core-collapse SNe, 200,000 type
Ia SNe and 20,000 luminous SNe per year.

\subsection{X-ray facilities}
By the time the SKA comes on line,  a number of telescopes very
relevant for SN studies should be operating \citep{Chandra15}.
These include the Cherenkov Telescope Array (CTA) (scheduled for 2019,
see Section~\ref{sec:CTA}), Chandra, NuSTAR, and hopefully Swift.
XMM-Newton has a planned end date in December 2018.
A planned X-ray telescope Athena+
(www.the-athena-x-ray-observatory.eu/) should be operational  in
2028 when both SKA and CTA are active.
There are several scientific synergies between SKA and CTA,  as made
clear from topics discussed in some recent joined meetings.\footnote{See, e.g., symposium 15 at  EWASS 2017,
http://eas.unige.ch/EWASS2017/.}
Athena+ will be sensitive in the energy band 0.2-12 keV and will
have an angular resolution of 5 arc-second at an energy $<$ 8 keV.
Sub-arcsecond angular resolution X-ray imaging combined with a
sensitive high resolution spectrometry of the proposed {\sl Lynx
X-ray Observatory} (https://wwwastro.msfc.nasa.gov/lynx/) are very
promising for the deep studies of both SNe and SNRs.

Until very recently no type Ia SN explosion has been confirmed in
X-rays, but \citet{Bochenek18}  have reported the detection in
Chandra data of SN 2012c, a type Ia-CSM object, i.e., a type Ia SN
surrounded by very dense material. On the other hand more than 60
core-collapse SNe have been detected in  X-rays \citep[see][and the
SNaX database at http://kronos.uchicago.edu/snax/]{Ross17}.

The thermal X-ray continuum  (i.e., \brems\ radiation) and the
thermal X-ray line emission of the shocked and heated CSM and ejecta
material behind the forward and reverse shocks respectively, are
tracers of the total density of the material into which the shocks
are expanding as well as the shocked temperature.
The evolution of the X-ray lightcurve can be  used to understand the
density structure of the material, and hence the mass-loss from the
progenitor star (see Section~\ref{sec:CSM} and the discussion in the
referenced papers).
The nonthermal radio and X-ray  \syn\ emission from \rel\ electrons
generated by the forward and reverse shocks, can strongly constrain
the shock acceleration process including the electron acceleration
efficiency and the magnetic field strength in the acceleration zone.
As described in Section~\ref{sec:broad}, modeling  the thermal and
nonthermal radiation consistently can constrain the overall \Facc\
efficiency and the electron-to-proton ratio.

\subsection{The Cherenkov Telescope Array} \label{sec:CTA}
The {\sl Cherenkov Telescope Array} (CTA, www.cta-observatory.org/)
is a \gamray\ observatory complex scheduled for completion in 2019.
It will be located in the Canary islands and in Chile. The northern site  will focus on low- and mid-energy ranges from 20 GeV to 20 TeV.
The southern site will also include some large telescopes sensitive in the high-energy band up to 100 TeV.
The expected \gamray\ sensitivity is on the order of 0.2\% Crab in 50 hours of observation at 1 TeV.  At energies above 100 TeV, LHAASO
(located in Tibet) is expected to reach a sensitivity of 0.01 Crab in 1 year \citep{Disciascio16}.

Extending the energy range of sensitive \gamray\ telescopes to
10 TeV and beyond is critical for testing CR acceleration at fast shocks. As mentioned in Section~\ref{sec:broad}, \IC\ from leptons competes with
\pion\ emission from hadrons at energies
up to $\sim 10$\,TeV, and which process dominates depends on often poorly constrained environmental factors. For higher electron energies however, IC emission drops due to the Klein-Nishina effect. Gamma-rays detected well above 10 TeV will almost  certainly be of hadronic origin. Such a discovery would constitute a breakthrough in CR physics proving that the hadronic CR component can
be accelerated quite early in the lifetime of a
SNR \citep{SB13}.  From calculations  of
CR production performed in \cite{Marcowith14} using a nonlinear \Facc\ model including MFA, the horizon for \gamray\ detection is
typically 10-20 Mpc.

The detection rate of extragalactic SNe in \gamrays\ is uncertain since no \gamray\ signal has yet been detected from such objects. Upper limits been reported \citep[i.e.,][]{Simoni17} and the expected number of SNe to be observed by CTA is strongly model dependent and highly uncertain.
However, CTA, together with SKA, LSST and some X-ray facilities will be active at the same time. This highlights the importance of using multi-messenger radio, optical and X-ray wave bands as triggers for \gamray\  monitoring for luminous sources  close enough not to be absorbed by the cosmic microwave background.

\section{Summary} \label{sec:sum}
Apart from the limited number of observed astrophysical neutrinos, and the small amount of interstellar dust so  far collected, CRs are the only source of mass from beyond the solar system.
Because of this, CRs convey information not obtainable
from photons, and understanding their origin is of fundamental importance for astrophysics.

As we emphasize in this review, supernovae, both extragalactic,
where the explosion and its aftereffect can often be witnessed, and
galactic remnants, which can be studied in detail from radio to
\gamrays, offer the best way to study the {\it in situ} acceleration
of CRs. While other sources contribute to the CRs observed at Earth
at some level, SNRs are clearly the source of the bulk of CRs with
energies below the knee at $\sim 10^{15}$\,eV
(Section~\ref{sec:intro}).
As for the acceleration mechanism, \FoFSA\ is  the most highly
developed theoretically, and has by far the most observational
confirmation, of any proposed mechanism. While other mechanisms
(e.g., those occurring in pulsars) may be important at some
level, \Facc\ in collisionless shocks is most likely the prime
accelerator of CRs and we concentrate on some of the current
problems facing nonlinear models.

The origin of CRs with energies beyond the knee  is far more
uncertain than those below the knee. We discuss the possibilities of
CR acceleration well beyond PeV energies by galactic sources
associated with the infrequent but powerful relativistic supernovae
which can accelerate CRs up to EeV (Section~\ref{S:PEV}). In
addition, we show how superluminous interaction powered supernovae,
and normal core-collapsed SNe in clusters of young massive stars,
can efficiently accelerate CRs to above PeV. At some energy above $
10^{18}$\,eV CRs are almost certainly of extragalactic origin  and
we mention some possible acceleration scenarios for such high
energies. A critical piece of the puzzle is the transition between
galactic and extragalactic CRs and understanding the production of
galactic CRs in SNRs is essential for modeling this transition
region.

While some aspects of CR origin and \Facc\ remain uncertain, we have
concentrated on \MFA, \Facc\ in \transrel\ shocks, acceleration in
star clusters from multiple shocks, and prospects for future
multi-messenger observations from radio to \gamray\ energies.

\begin{acknowledgements}
 A.M.B, D.C.E and A.M thank the staff of ISSI for their generous
hospitality and assistance. The authors thank the referees for the
constructive comments. A.M.~Bykov and S.M.~Osipov were supported by
the RSF grant 16-12-10225. Some of the modeling was performed at the
``Tornado'' subsystem of the St.~Petersburg Polytechnic University
supercomputing center. A.M.~Bykov thanks R.A.~Chevalier and
J.C.~Raymond for discussions, R.Margutti for Figure 1, and
M.A.~Grekov for his support with computations.
\end{acknowledgements}

\bibliographystyle{aa} 
\bibliography{c:/a_a_TOP/bibTeX/bib_DCE_Alex}

\begin{thebibliography}{212}
\expandafter\ifx\csname natexlab\endcsname\relax\def\natexlab#1{#1}\fi

\bibitem[{{Aartsen} {et~al.}(2014){Aartsen}, {Ackermann}, {Adams}, {Aguilar},
  {Ahlers}, {Ahrens}, {Altmann}, {Anderson}, {Arguelles}, {Arlen}, \&
  et~al.}]{Aartsen14}
{Aartsen}, M.~G., {Ackermann}, M., {Adams}, J., {et~al.} 2014, Physical Review
  Letters, 113, 101101

\bibitem[{{Abdo} {et~al.}(2010{\natexlab{a}}){Abdo}, {Ackermann}, {Ajello},
  {Baldini}, {Ballet}, {Barbiellini}, {Baring}, {Bastieri}, {Baughman},
  {Bechtol}, {Bellazzini}, {Berenji}, {Blandford}, {Bloom}, {Bonamente},
  {Borgland}, {Bregeon}, {Brez}, {Brigida}, {Bruel}, {Burnett}, {Buson},
  {Caliandro}, {Cameron}, {Caraveo}, {Casandjian}, {Cecchi}, {{\c C}elik},
  {Chekhtman}, {Cheung}, {Chiang}, {Ciprini}, {Claus}, {Cognard},
  {Cohen-Tanugi}, {Cominsky}, {Conrad}, {Cutini}, {Dermer}, {de Angelis}, {de
  Palma}, {Digel}, {do Couto e Silva}, {Drell}, {Dubois}, {Dumora}, {Espinoza},
  {Farnier}, {Favuzzi}, {Fegan}, {Focke}, {Fortin}, {Frailis}, {Fukazawa},
  {Funk}, {Fusco}, {Gargano}, {Gasparrini}, {Gehrels}, {Germani}, {Giavitto},
  {Giebels}, {Giglietto}, {Giordano}, {Glanzman}, {Godfrey}, {Grenier},
  {Grondin}, {Grove}, {Guillemot}, {Guiriec}, {Hanabata}, {Harding},
  {Hayashida}, {Hays}, {Hughes}, {Jackson}, {J{\'o}hannesson}, {Johnson},
  {Johnson}, {Johnson}, {Kamae}, {Katagiri}, {Kataoka}, {Katsuta}, {Kawai},
  {Kerr}, {Kn{\"o}dlseder}, {Kocian}, {Kramer}, {Kuss}, {Lande}, {Latronico},
  {Lemoine-Goumard}, {Longo}, {Loparco}, {Lott}, {Lovellette}, {Lubrano},
  {Lyne}, {Madejski}, {Makeev}, {Mazziotta}, {McEnery}, {Meurer}, {Michelson},
  {Mitthumsiri}, {Mizuno}, {Monte}, {Monzani}, {Morselli}, {Moskalenko},
  {Murgia}, {Nakamori}, {Nolan}, {Norris}, {Noutsos}, {Nuss}, {Ohsugi},
  {Omodei}, {Orlando}, {Ormes}, {Paneque}, {Parent}, {Pelassa}, {Pepe},
  {Pesce-Rollins}, {Piron}, {Porter}, {Rain{\`o}}, {Rando}, {Razzano},
  {Reimer}, {Reimer}, {Reposeur}, {Rochester}, {Rodriguez}, {Romani}, {Roth},
  {Ryde}, {Sadrozinski}, {Sanchez}, {Sander}, {Parkinson}, {Scargle},
  {Sgr{\`o}}, {Siskind}, {Smith}, {Smith}, {Spandre}, {Spinelli}, {Stappers},
  {Stecker}, {Strickman}, {Suson}, {Tajima}, {Takahashi}, {Takahashi},
  {Tanaka}, {Thayer}, {Thayer}, {Theureau}, {Thompson}, {Tibaldo}, {Tibolla},
  {Torres}, {Tosti}, {Tramacere}, {Uchiyama}, {Usher}, {Vasileiou}, {Venter},
  {Vilchez}, {Vitale}, {Waite}, {Wang}, {Winer}, {Wood}, {Yamazaki}, {Ylinen},
  \& {Ziegler}}]{AbdoEtalW442010}
{Abdo}, A.~A., {Ackermann}, M., {Ajello}, M., {et~al.} 2010{\natexlab{a}},
  Science, 327, 1103

\bibitem[{{Abdo} {et~al.}(2010{\natexlab{b}}){Abdo}, {Ackermann}, {Ajello},
  {Baldini}, {Ballet}, {Barbiellini}, {Bastieri}, {Baughman}, {Bechtol},
  {Bellazzini}, {Berenji}, {Blandford}, {Bloom}, {Bonamente}, {Borgland},
  {Bregeon}, {Brez}, {Brigida}, {Bruel}, {Burnett}, {Buson}, {Caliandro},
  {Cameron}, {Caraveo}, {Casandjian}, {Cecchi}, {{\c C}elik}, {Chekhtman},
  {Cheung}, {Chiang}, {Cillis}, {Ciprini}, {Claus}, {Cohen-Tanugi}, {Cominsky},
  {Conrad}, {Cutini}, {Dermer}, {de Angelis}, {de Palma}, {Silva}, {Drell},
  {Drlica-Wagner}, {Dubois}, {Dumora}, {Farnier}, {Favuzzi}, {Fegan}, {Focke},
  {Fortin}, {Frailis}, {Fukazawa}, {Funk}, {Fusco}, {Gargano}, {Gasparrini},
  {Gehrels}, {Germani}, {Giavitto}, {Giebels}, {Giglietto}, {Giordano},
  {Glanzman}, {Godfrey}, {Grenier}, {Grondin}, {Grove}, {Guillemot}, {Guiriec},
  {Hanabata}, {Harding}, {Hayashida}, {Hughes}, {Jackson}, {J{\'o}hannesson},
  {Johnson}, {Johnson}, {Johnson}, {Kamae}, {Katagiri}, {Kataoka}, {Kawai},
  {Kerr}, {Kn{\"o}dlseder}, {Kocian}, {Kuss}, {Lande}, {Latronico}, {Lee},
  {Lemoine-Goumard}, {Longo}, {Loparco}, {Lott}, {Lovellette}, {Lubrano},
  {Madejski}, {Makeev}, {Mazziotta}, {Meurer}, {Michelson}, {Mitthumsiri},
  {Moiseev}, {Monte}, {Monzani}, {Morselli}, {Moskalenko}, {Murgia},
  {Nakamori}, {Nolan}, {Norris}, {Nuss}, {Ohsugi}, {Orlando}, {Ormes}, {Ozaki},
  {Paneque}, {Panetta}, {Parent}, {Pelassa}, {Pepe}, {Pesce-Rollins}, {Piron},
  {Porter}, {Rain{\`o}}, {Rando}, {Razzano}, {Reimer}, {Reimer}, {Reposeur},
  {Rochester}, {Rodriguez}, {Romani}, {Roth}, {Ryde}, {Sadrozinski}, {Sanchez},
  {Sander}, {Saz Parkinson}, {Scargle}, {Sgr{\`o}}, {Siskind}, {Smith},
  {Smith}, {Spandre}, {Spinelli}, {Strickman}, {Strong}, {Suson}, {Tajima},
  {Takahashi}, {Takahashi}, {Tanaka}, {Thayer}, {Thayer}, {Thompson},
  {Tibaldo}, {Torres}, {Tosti}, {Tramacere}, {Uchiyama}, {Usher}, {Van Etten},
  {Vasileiou}, {Venter}, {Vilchez}, {Vitale}, {Waite}, {Wang}, {Winer}, {Wood},
  {Ylinen}, \& {Ziegler}}]{AbdoEtalIC4432010}
{Abdo}, A.~A., {Ackermann}, M., {Ajello}, M., {et~al.} 2010{\natexlab{b}},
  \apj, 712, 459

\bibitem[{{Abeysekara} {et~al.}(2017){Abeysekara}, {Albert}, {Alfaro},
  {Alvarez}, {{\'A}lvarez}, {Arceo}, {Arteaga-Vel{\'a}zquez}, {Avila Rojas},
  {Ayala Solares}, {Barber}, {Becerra Gonzalez}, {Becerril}, {Belmont-Moreno},
  {BenZvi}, {Berley}, {Bernal}, {Brisbois}, {Caballero-Mora}, {Capistr{\'a}n},
  {Carrami{\~n}ana}, {Casanova}, {Castillo}, {Cotti}, {Cotzomi}, {Couti{\~n}o
  de Le{\'o}n}, {De la Fuente}, {De Le{\'o}n}, {DeYoung}, {Diaz Hernandez},
  {Diaz-Cruz}, {D{\'{\i}}az-V{\'e}lez}, {Dichiara}, {Dingus}, {DuVernois},
  {Ellsworth}, {Engel}, {Enriquez-Rivera}, {Fick}, {Fiorino}, {Fleischhack},
  {Flores}, {Fraija}, {Garc{\'{\i}}a-Gonz{\'a}lez}, {Garcia-Luna},
  {Garcia-Torales}, {Garfias}, {Gerhardt}, {Gonz{\'a}lez}, {Gonz{\'a}lez
  Mu{\~n}oz}, {Goodman}, {Gussert}, {Hampel-Arias}, {Harding}, {Hernandez},
  {Hernandez-Almada}, {Hinton}, {Hona}, {Hui}, {H{\"u}ntemeyer}, {Iriarte},
  {Jardin-Blicq}, {Joshi}, {Kaufmann}, {Kieda}, {Kunde}, {Lara}, {Lauer},
  {Lee}, {Lennarz}, {Le{\'o}n Vargas}, {Linnemann}, {Longinotti}, {Longo
  Proper}, {L{\'o}pez-Coto}, {Raya}, {Luna-Garc{\'{\i}}a}, {Malone},
  {Marandon}, {Marinelli}, {Martinez}, {Martinez-Castellanos},
  {Mart{\'{\i}}nez-Castro}, {Mart{\'{\i}}nez-Huerta}, {Matthews}, {McEnery},
  {Miranda-Romagnoli}, {Moreno}, {Mostaf{\'a}}, {Nellen}, {Newbold}, {Nisa},
  {Noriega-Papaqui}, {Pelayo}, {P{\'e}rez-P{\'e}rez}, {Pretz}, {Ren}, {Rho},
  {Rivi{\`e}re}, {Rosa-Gonz{\'a}lez}, {Rosenberg}, {Ruiz-Velasco}, {Ryan},
  {Salazar}, {Salesa Greus}, {Sandoval}, {Schneider}, {Schoorlemmer}, {Sinnis},
  {Smith}, {Smith}, {Springer}, {Surajbali}, {Taboada}, {Tibolla}, {Tollefson},
  {Torres}, {Ukwatta}, {Vianello}, {Weisgarber}, {Westerhoff}, {Wood},
  {Yapici}, {Yodh}, {Younk}, {Zepeda}, \& {Zhou}}]{HAWC_35ICRC}
{Abeysekara}, A.~U., {Albert}, A., {Alfaro}, R., {et~al.} 2017, ArXiv e-prints

\bibitem[{{Acero} {et~al.}(2016){Acero}, {Ackermann}, {Ajello}, {Baldini},
  {Ballet}, {Barbiellini}, {Bastieri}, {Bellazzini}, {Bissaldi}, {Blandford},
  {Bloom}, {Bonino}, {Bottacini}, {Brandt}, {Bregeon}, {Bruel}, {Buehler},
  {Buson}, {Caliandro}, {Cameron}, {Caputo}, {Caragiulo}, {Caraveo},
  {Casandjian}, {Cavazzuti}, {Cecchi}, {Chekhtman}, {Chiang}, {Chiaro},
  {Ciprini}, {Claus}, {Cohen}, {Cohen-Tanugi}, {Cominsky}, {Condon}, {Conrad},
  {Cutini}, {D'Ammando}, {de Angelis}, {de Palma}, {Desiante}, {Digel}, {Di
  Venere}, {Drell}, {Drlica-Wagner}, {Favuzzi}, {Ferrara}, {Franckowiak},
  {Fukazawa}, {Funk}, {Fusco}, {Gargano}, {Gasparrini}, {Giglietto}, {Giommi},
  {Giordano}, {Giroletti}, {Glanzman}, {Godfrey}, {Gomez-Vargas}, {Grenier},
  {Grondin}, {Guillemot}, {Guiriec}, {Gustafsson}, {Hadasch}, {Harding},
  {Hayashida}, {Hays}, {Hewitt}, {Hill}, {Horan}, {Hou}, {Iafrate}, {Jogler},
  {J{\'o}hannesson}, {Johnson}, {Kamae}, {Katagiri}, {Kataoka}, {Katsuta},
  {Kerr}, {Kn{\"o}dlseder}, {Kocevski}, {Kuss}, {Laffon}, {Lande}, {Larsson},
  {Latronico}, {Lemoine-Goumard}, {Li}, {Li}, {Longo}, {Loparco}, {Lovellette},
  {Lubrano}, {Magill}, {Maldera}, {Marelli}, {Mayer}, {Mazziotta}, {Michelson},
  {Mitthumsiri}, {Mizuno}, {Moiseev}, {Monzani}, {Moretti}, {Morselli},
  {Moskalenko}, {Murgia}, {Nemmen}, {Nuss}, {Ohsugi}, {Omodei}, {Orienti},
  {Orlando}, {Ormes}, {Paneque}, {Perkins}, {Pesce-Rollins}, {Petrosian},
  {Piron}, {Pivato}, {Porter}, {Rain{\`o}}, {Rando}, {Razzano}, {Razzaque},
  {Reimer}, {Reimer}, {Renaud}, {Reposeur}, {Rousseau}, {Saz Parkinson},
  {Schmid}, {Schulz}, {Sgr{\`o}}, {Siskind}, {Spada}, {Spandre}, {Spinelli},
  {Strong}, {Suson}, {Tajima}, {Takahashi}, {Tanaka}, {Thayer}, {Thompson},
  {Tibaldo}, {Tibolla}, {Torres}, {Tosti}, {Troja}, {Uchiyama}, {Vianello},
  {Wells}, {Wood}, {Wood}, {Yassine}, {den Hartog}, \& {Zimmer}}]{Acero16}
{Acero}, F., {Ackermann}, M., {Ajello}, M., {et~al.} 2016, \apjs, 224, 8

\bibitem[{{Acero} {et~al.}(2015){Acero}, {Lemoine-Goumard}, {Renaud}, {Ballet},
  {Hewitt}, {Rousseau}, \& {Tanaka}}]{Acero15}
{Acero}, F., {Lemoine-Goumard}, M., {Renaud}, M., {et~al.} 2015, \aap, 580, A74

\bibitem[{{Achterberg} {et~al.}(1994){Achterberg}, {Blandford}, \&
  {Reynolds}}]{Achterberg94}
{Achterberg}, A., {Blandford}, R.~D., \& {Reynolds}, S.~P. 1994, \aap, 281, 220

\bibitem[{{Ackermann} {et~al.}(2013{\natexlab{a}}){Ackermann}, {Ajello},
  {Allafort}, {Baldini}, {Ballet}, {Barbiellini}, {Baring}, {Bastieri},
  {Bechtol}, {Bellazzini}, {Blandford}, {Bloom}, {Bonamente}, {Borgland},
  {Bottacini}, {Brandt}, {Bregeon}, {Brigida}, {Bruel}, {Buehler}, {Busetto},
  {Buson}, {Caliandro}, {Cameron}, {Caraveo}, {Casandjian}, {Cecchi}, {{\c
  C}elik}, {Charles}, {Chaty}, {Chaves}, {Chekhtman}, {Cheung}, {Chiang},
  {Chiaro}, {Cillis}, {Ciprini}, {Claus}, {Cohen-Tanugi}, {Cominsky}, {Conrad},
  {Corbel}, {Cutini}, {D'Ammando}, {de Angelis}, {de Palma}, {Dermer}, {do
  Couto e Silva}, {Drell}, {Drlica-Wagner}, {Falletti}, {Favuzzi}, {Ferrara},
  {Franckowiak}, {Fukazawa}, {Funk}, {Fusco}, {Gargano}, {Germani},
  {Giglietto}, {Giommi}, {Giordano}, {Giroletti}, {Glanzman}, {Godfrey},
  {Grenier}, {Grondin}, {Grove}, {Guiriec}, {Hadasch}, {Hanabata}, {Harding},
  {Hayashida}, {Hayashi}, {Hays}, {Hewitt}, {Hill}, {Hughes}, {Jackson},
  {Jogler}, {J{\'o}hannesson}, {Johnson}, {Kamae}, {Kataoka}, {Katsuta},
  {Kn{\"o}dlseder}, {Kuss}, {Lande}, {Larsson}, {Latronico}, {Lemoine-Goumard},
  {Longo}, {Loparco}, {Lovellette}, {Lubrano}, {Madejski}, {Massaro}, {Mayer},
  {Mazziotta}, {McEnery}, {Mehault}, {Michelson}, {Mignani}, {Mitthumsiri},
  {Mizuno}, {Moiseev}, {Monzani}, {Morselli}, {Moskalenko}, {Murgia},
  {Nakamori}, {Nemmen}, {Nuss}, {Ohno}, {Ohsugi}, {Omodei}, {Orienti},
  {Orlando}, {Ormes}, {Paneque}, {Perkins}, {Pesce-Rollins}, {Piron}, {Pivato},
  {Rain{\`o}}, {Rando}, {Razzano}, {Razzaque}, {Reimer}, {Reimer}, {Ritz},
  {Romoli}, {S{\'a}nchez-Conde}, {Schulz}, {Sgr{\`o}}, {Simeon}, {Siskind},
  {Smith}, {Spandre}, {Spinelli}, {Stecker}, {Strong}, {Suson}, {Tajima},
  {Takahashi}, {Takahashi}, {Tanaka}, {Thayer}, {Thayer}, {Thompson},
  {Thorsett}, {Tibaldo}, {Tibolla}, {Tinivella}, {Troja}, {Uchiyama}, {Usher},
  {Vandenbroucke}, {Vasileiou}, {Vianello}, {Vitale}, {Waite}, {Werner},
  {Winer}, {Wood}, {Wood}, {Yamazaki}, {Yang}, \& {Zimmer}}]{AckermannSci13}
{Ackermann}, M., {Ajello}, M., {Allafort}, A., {et~al.} 2013{\natexlab{a}},
  Science, 339, 807

\bibitem[{{Ackermann} {et~al.}(2013{\natexlab{b}}){Ackermann}, {Ajello},
  {Allafort}, {Baldini}, {Ballet}, {Barbiellini}, {Baring}, {Bastieri},
  {Bechtol}, {Bellazzini}, {Blandford}, \& et~al.}]{AckermannEtal_W44}
{Ackermann}, M., {Ajello}, M., {Allafort}, A., {et~al.} 2013{\natexlab{b}},
  Science, 339, 807

\bibitem[{{Ackermann} {et~al.}(2011){Ackermann}, {Ajello}, {Allafort},
  {Baldini}, {Ballet}, {Barbiellini}, {Bastieri}, {Belfiore}, {Bellazzini},
  {Berenji}, {Blandford}, {Bloom}, {Bonamente}, {Borgland}, {Bottacini},
  {Brigida}, {Bruel}, {Buehler}, {Buson}, {Caliandro}, {Cameron}, {Caraveo},
  {Casandjian}, {Cecchi}, {Chekhtman}, {Cheung}, {Chiang}, {Ciprini}, {Claus},
  {Cohen-Tanugi}, {de Angelis}, {de Palma}, {Dermer}, {do Couto e Silva},
  {Drell}, {Dumora}, {Favuzzi}, {Fegan}, {Focke}, {Fortin}, {Fukazawa},
  {Fusco}, {Gargano}, {Germani}, {Giglietto}, {Giordano}, {Giroletti},
  {Glanzman}, {Godfrey}, {Grenier}, {Guillemot}, {Guiriec}, {Hadasch},
  {Hanabata}, {Okumura}, {Orlando}, {Ormes}, {Ozaki}, {Paneque}, {Parent},
  {Pesce-Rollins}, {Pierbattista}, {Piron}, {Pohl}, {Prokhorov}, {Rain{\`o}},
  {Rando}, {Razzano}, {Reposeur}, {Ritz}, {Parkinson}, {Sgr{\`o}}, {Siskind},
  {Smith}, {Spinelli}, {Strong}, {Takahashi}, {Tanaka}, {Thayer}, {Thayer},
  {Thompson}, {Tibaldo}, {Torres}, {Tosti}, {Tramacere}, {Troja}, {Uchiyama},
  {Vandenbroucke}, {Vasileiou}, {Vianello}, {Vitale}, {Waite}, {Wang}, {Winer},
  {Wood}, {Yang}, {Zimmer}, \& {Bontemps}}]{AckermannSB2011}
{Ackermann}, M., {Ajello}, M., {Allafort}, A., {et~al.} 2011, Science, 334,
  1103

\bibitem[{{Aharonian} {et~al.}(2002){Aharonian}, {Akhperjanian}, {Beilicke},
  {Bernl{\"o}hr}, {B{\"o}rst}, {Bojahr}, {Bolz}, {Coarasa}, {Contreras},
  {Cortina}, {Denninghoff}, {Fonseca}, {Girma}, {G{\"o}tting}, {Heinzelmann},
  {Hermann}, {Heusler}, {Hofmann}, {Horns}, {Jung}, {Kankanyan}, {Kestel},
  {Kettler}, {Kohnle}, {Konopelko}, {Kornmeyer}, {Kranich}, {Krawczynski},
  {Lampeitl}, {Lopez}, {Lorenz}, {Lucarelli}, {Magnussen}, {Mang}, {Meyer},
  {Milite}, {Mirzoyan}, {Moralejo}, {Ona}, {Panter}, {Plyasheshnikov}, {Prahl},
  {P{\"u}hlhofer}, {Rauterberg}, {Reyes}, {Rhode}, {Ripken}, {R{\"o}hring},
  {Rowell}, {Sahakian}, {Samorski}, {Schilling}, {Schr{\"o}der}, {Siems},
  {Sobzynska}, {Stamm}, {Tluczykont}, {V{\"o}lk}, {Wiedner}, {Wittek},
  {Uchiyama}, {Takahashi}, \& {HEGRA Collaboration}}]{AharonianEtal2002}
{Aharonian}, F., {Akhperjanian}, A., {Beilicke}, M., {et~al.} 2002, \aap, 393,
  L37

\bibitem[{{Aharonian} {et~al.}(2012){Aharonian}, {Bykov}, {Parizot}, {Ptuskin},
  \& {Watson}}]{ABPW2012}
{Aharonian}, F., {Bykov}, A., {Parizot}, E., {Ptuskin}, V., \& {Watson}, A.
  2012, \ssr, 166, 97

\bibitem[{{Aharonian} \& {Neronov}(2005)}]{Aharonian2005}
{Aharonian}, F. \& {Neronov}, A. 2005, \apj, 619, 306

\bibitem[{{Ahn} {et~al.}(2010){Ahn}, {Allison}, {Bagliesi}, {Beatty},
  {Bigongiari}, {Childers}, {Conklin}, {Coutu}, {DuVernois}, {Ganel}, {Han},
  {Jeon}, {Kim}, {Lee}, {Lutz}, {Maestro}, {Malinin}, {Marrocchesi}, {Minnick},
  {Mognet}, {Nam}, {Nam}, {Nutter}, {Park}, {Park}, {Seo}, {Sina}, {Wu},
  {Yang}, {Yoon}, {Zei}, \& {Zinn}}]{AhnEtal2010}
{Ahn}, H.~S., {Allison}, P., {Bagliesi}, M.~G., {et~al.} 2010, \apjl, 714, L89

\bibitem[{{Ahnen} {et~al.}(2017){Ahnen}, {Ansoldi}, {Antonelli}, {Arcaro},
  {Babi{\'c}}, {Banerjee}, {Bangale}, {Barres de Almeida}, {Barrio}, {Becerra
  Gonz{\'a}lez}, {Bednarek}, {Bernardini}, {Berti}, {Bhattacharyya},
  {Biasuzzi}, {Biland}, {Blanch}, {Bonnefoy}, {Bonnoli}, {Carosi}, {Carosi},
  {Chatterjee}, {Colak}, {Colin}, {Colombo}, {Contreras}, {Cortina}, {Covino},
  {Cumani}, {Da Vela}, {Dazzi}, {De Angelis}, {De Lotto}, {de O{\~n}a
  Wilhelmi}, {Di Pierro}, {Doert}, {Dom{\'{\i}}nguez}, {Dominis Prester},
  {Dorner}, {Doro}, {Einecke}, {Eisenacher Glawion}, {Elsaesser},
  {Engelkemeier}, {Fallah Ramazani}, {Fern{\'a}ndez-Barral}, {Fidalgo},
  {Fonseca}, {Font}, {Fruck}, {Galindo}, {Garc{\'{\i}}a L{\'o}pez},
  {Garczarczyk}, {Gaug}, {Giammaria}, {Godinovi{\'c}}, {Gora}, {Guberman},
  {Hadasch}, {Hahn}, {Hassan}, {Hayashida}, {Herrera}, {Hose}, {Hrupec},
  {Inada}, {Ishio}, {Konno}, {Kubo}, {Kushida}, {Kuve{\v z}di{\'c}}, {Lelas},
  {Lindfors}, {Lombardi}, {Longo}, {L{\'o}pez}, {Maggio}, {Majumdar},
  {Makariev}, {Maneva}, {Manganaro}, {Mannheim}, {Maraschi}, {Mariotti},
  {Mart{\'{\i}}nez}, {Mazin}, {Menzel}, {Minev}, {Mirzoyan}, {Moralejo},
  {Moreno}, {Moretti}, {Neustroev}, {Niedzwiecki}, {Nievas Rosillo}, {Nilsson},
  {Ninci}, {Nishijima}, {Noda}, {Nogu{\'e}s}, {Paiano}, {Palacio}, {Paneque},
  {Paoletti}, {Paredes}, {Pedaletti}, {Peresano}, {Perri}, {Persic}, {Prada
  Moroni}, {Prandini}, {Puljak}, {Garcia}, {Reichardt}, {Rhode}, {Rib{\'o}},
  {Rico}, {Righi}, {Saito}, {Satalecka}, {Schroeder}, {Schweizer}, {Shore},
  {Sitarek}, {{\v S}nidari{\'c}}, {Sobczynska}, {Stamerra}, {Strzys},
  {Suri{\'c}}, {Takalo}, {Tavecchio}, {Temnikov}, {Terzi{\'c}}, {Tescaro},
  {Teshima}, {Torres-Alb{\`a}}, {Treves}, {Vanzo}, {Vazquez Acosta}, {Vovk},
  {Ward}, {Will}, \& {Zari{\'c}}}]{MAGIC_CasA2017}
{Ahnen}, M.~L., {Ansoldi}, S., {Antonelli}, L.~A., {et~al.} 2017, \mnras, 472,
  2956

\bibitem[{{Alexander} {et~al.}(2015){Alexander}, {Soderberg}, \&
  {Chomiuk}}]{Alexander15}
{Alexander}, K.~D., {Soderberg}, A.~M., \& {Chomiuk}, L.~B. 2015, \apj, 806,
  106

\bibitem[{{Aloisio} {et~al.}(2007){Aloisio}, {Berezinsky}, {Blasi}, {Gazizov},
  {Grigorieva}, \& {Hnatyk}}]{AloisioEtal2007}
{Aloisio}, R., {Berezinsky}, V., {Blasi}, P., {et~al.} 2007, Astroparticle
  Physics, 27, 76

\bibitem[{{Amato}(2014)}]{amato14}
{Amato}, E. 2014, International Journal of Modern Physics D, 23, 30013

\bibitem[{{Apel} {et~al.}(2013){Apel}, {Arteaga-Vel{\`a}zquez}, {Bekk},
  {Bertaina}, {Bl{\"u}mer}, {Bozdog}, {Brancus}, {Cantoni}, {Chiavassa},
  {Cossavella}, {Daumiller}, {de Souza}, {Di Pierro}, {Doll}, {Engel},
  {Engler}, {Finger}, {Fuchs}, {Fuhrmann}, {Gils}, {Glasstetter}, {Grupen},
  {Haungs}, {Heck}, {H{\"o}randel}, {Huber}, {Huege}, {Kampert}, {Kang},
  {Klages}, {Link}, {{\L}uczak}, {Ludwig}, {Mathes}, {Mayer}, {Melissas},
  {Milke}, {Mitrica}, {Morello}, {Oehlschl{\"a}ger}, {Ostapchenko}, {Palmieri},
  {Petcu}, {Pierog}, {Rebel}, {Roth}, {Schieler}, {Schoo}, {Schr{\"o}der},
  {Sima}, {Toma}, {Trinchero}, {Ulrich}, {Weindl}, {Wochele}, {Wommer}, \&
  {Zabierowski}}]{ApelEtal2013}
{Apel}, W.~D., {Arteaga-Vel{\`a}zquez}, J.~C., {Bekk}, K., {et~al.} 2013, \prd,
  87, 081101

\bibitem[{{Archambault} {et~al.}(2017){Archambault}, {Archer}, {Benbow},
  {Bird}, {Bourbeau}, {Buchovecky}, {Buckley}, {Bugaev}, {Cerruti}, {Connolly},
  {Cui}, {Dwarkadas}, {Errando}, {Falcone}, {Feng}, {Finley}, {Fleischhack},
  {Fortson}, {Furniss}, {Griffin}, {H{\"u}tten}, {Hanna}, {Holder}, {Johnson},
  {Kaaret}, {Kar}, {Kelley-Hoskins}, {Kertzman}, {Kieda}, {Krause}, {Kumar},
  {Lang}, {Maier}, {McArthur}, {McCann}, {Moriarty}, {Mukherjee}, {Nieto},
  {O'Brien}, {Ong}, {Otte}, {Park}, {Pohl}, {Popkow}, {Pueschel}, {Quinn},
  {Ragan}, {Reynolds}, {Richards}, {Roache}, {Sadeh}, {Santander}, {Sembroski},
  {Shahinyan}, {Slane}, {Staszak}, {Telezhinsky}, {Trepanier}, {Tyler},
  {Wakely}, {Weinstein}, {Weisgarber}, {Wilcox}, {Wilhelm}, {Williams}, \&
  {Zitzer}}]{Archambault2017}
{Archambault}, S., {Archer}, A., {Benbow}, W., {et~al.} 2017, \apj, 836, 23

\bibitem[{{Arons}(2003)}]{Arons2003}
{Arons}, J. 2003, \apj, 589, 871

\bibitem[{{Asano} \& {M{\'e}sz{\'a}ros}(2016)}]{Asano2016}
{Asano}, K. \& {M{\'e}sz{\'a}ros}, P. 2016, \prd, 94, 023005

\bibitem[{{Axford}(1981)}]{Axford81}
{Axford}, W.~I. 1981, in IAU Symposium, Vol.~94, Origin of Cosmic Rays, ed.
  G.~{Setti}, G.~{Spada}, \& A.~W. {Wolfendale}, 339--358

\bibitem[{{Axford} {et~al.}(1977){Axford}, {Leer}, \& {Skadron}}]{ALS77}
{Axford}, W.~I., {Leer}, E., \& {Skadron}, G. 1977, Proc. 15th ICRC(Plovdiv),
  11, 132

\bibitem[{Baade \& Zwicky(1934)}]{BZ34}
Baade, W. \& Zwicky, F. 1934, Phys. Rev., 46, 76

\bibitem[{{Ballet}(2006)}]{Ballet2006}
{Ballet}, J. 2006, Advances in Space Research, 37, 1902

\bibitem[{{Bartel} {et~al.}(2017){Bartel}, {Karimi}, \&
  {Bietenholz}}]{Bartel17}
{Bartel}, N., {Karimi}, B., \& {Bietenholz}, M.~F. 2017, Astronomy Reports, 61,
  299

\bibitem[{{Bell}(1978)}]{Bell78a}
{Bell}, A.~R. 1978, MNRAS, 182, 147

\bibitem[{{Bell}(2004)}]{Bell2004}
{Bell}, A.~R. 2004, \mnras, 353, 550

\bibitem[{{Bell}(2005)}]{Bell2005}
{Bell}, A.~R. 2005, \mnras, 358, 181

\bibitem[{{Bell} \& {Lucek}(2001)}]{bell_lucek01}
{Bell}, A.~R. \& {Lucek}, S.~G. 2001, \mnras, 321, 433

\bibitem[{{Bell} {et~al.}(2013){Bell}, {Schure}, {Reville}, \&
  {Giacinti}}]{BellEtal2013}
{Bell}, A.~R., {Schure}, K.~M., {Reville}, B., \& {Giacinti}, G. 2013, \mnras,
  431, 415

\bibitem[{Berezhko \& Ellison(1999)}]{BE99}
Berezhko, E.~G. \& Ellison, D.~C. 1999, ApJ, 526, 385

\bibitem[{{Berezhko} {et~al.}(1996){Berezhko}, {Elshin}, \&
  {Ksenofontov}}]{BEK96}
{Berezhko}, E.~G., {Elshin}, V.~K., \& {Ksenofontov}, L.~T. 1996, Soviet
  Journal of Experimental and Theoretical Physics, 82, 1

\bibitem[{{Berezhnev} {et~al.}(2012){Berezhnev}, {Besson}, {Budnev},
  {Chiavassa}, {Chvalaev}, {Gress}, {Dyachok}, {Epimakhov}, {Haungs}, {Karpov},
  {Kalmykov}, {Konstantinov}, {Korobchenko}, {Korosteleva}, {Kozhin},
  {Kuzmichev}, {Lubsandorzhiev}, {Lubsandorzhiev}, {Mirgazov}, {Panasyuk},
  {Pankov}, {Popova}, {Prosin}, {Ptuskin}, {Semeney}, {Shaibonov}, {Silaev},
  {Silaev}, {Skurikhin}, {Snyder}, {Spiering}, {Schr{\"o}der}, {Stockham},
  {Sveshnikova}, {Wischnewski}, {Yashin}, \& {Zagorodnikov}}]{berezhnevea13}
{Berezhnev}, S.~F., {Besson}, D., {Budnev}, N.~M., {et~al.} 2012, Nuclear
  Instruments and Methods in Physics Research A, 692, 98

\bibitem[{{Berezinskii} {et~al.}(1990){Berezinskii}, {Bulanov}, {Dogiel}, \&
  {Ptuskin}}]{Berezinski90}
{Berezinskii}, V.~S., {Bulanov}, S.~V., {Dogiel}, V.~A., \& {Ptuskin}, V.~S.
  1990, {Astrophysics of cosmic rays}

\bibitem[{{Bergman} \& {Belz}(2007)}]{Bergman2007}
{Bergman}, D.~R. \& {Belz}, J.~W. 2007, Journal of Physics G Nuclear Physics,
  34, R359

\bibitem[{{Bietenholz} {et~al.}(2010){Bietenholz}, {Bartel}, \&
  {Rupen}}]{Bietenholz10}
{Bietenholz}, M.~F., {Bartel}, N., \& {Rupen}, M.~P. 2010, \apj, 712, 1057

\bibitem[{{Binns} {et~al.}(2014){Binns}, {Christian}, {Cummings}, {de Nolfo},
  {Israel}, {Lave}, {Leske}, {Mewaldt}, {Stone}, {von Rosenvinge}, \&
  {Wiedenbeck}}]{BinnsEtal2014}
{Binns}, W.~R., {Christian}, E.~R., {Cummings}, A.~C., {et~al.} 2014, in APS
  April Meeting Abstracts

\bibitem[{{Binns} {et~al.}(2016){Binns}, {Israel}, {Christian}, {Cummings}, {de
  Nolfo}, {Lave}, {Leske}, {Mewaldt}, {Stone}, {von Rosenvinge}, \&
  {Wiedenbeck}}]{Binns_60Fe_2016}
{Binns}, W.~R., {Israel}, M.~H., {Christian}, E.~R., {et~al.} 2016, Science,
  352, 677

\bibitem[{{Bj{\"o}rnsson} \& {Keshavarzi}(2017)}]{Bjornsson17}
{Bj{\"o}rnsson}, C.-I. \& {Keshavarzi}, S.~T. 2017, \apj, 841, 12

\bibitem[{{Blandford} \& {Eichler}(1987)}]{be87}
{Blandford}, R. \& {Eichler}, D. 1987, Physics Reports, 154, 1

\bibitem[{{Blandford} {et~al.}(2014){Blandford}, {Simeon}, \&
  {Yuan}}]{blandford14}
{Blandford}, R., {Simeon}, P., \& {Yuan}, Y. 2014, Nuclear Physics B
  Proceedings Supplements, 256, 9

\bibitem[{{Blandford} \& {Ostriker}(1978)}]{BO78}
{Blandford}, R.~D. \& {Ostriker}, J.~P. 1978, ApJ, 221, L29

\bibitem[{{Blasi}(2013)}]{blasi2013}
{Blasi}, P. 2013, \aapr, 21, 70

\bibitem[{{Blasi} {et~al.}(2000){Blasi}, {Epstein}, \&
  {Olinto}}]{blasi_magnetar_2000}
{Blasi}, P., {Epstein}, R.~I., \& {Olinto}, A.~V. 2000, \apjl, 533, L123

\bibitem[{{Blinnikov}(2016{\natexlab{a}})}]{Blinnikov2016B}
{Blinnikov}, S. 2016{\natexlab{a}}, ArXiv e-prints 1611.00513

\bibitem[{{Blinnikov}(2016{\natexlab{b}})}]{Blinnikov2016}
{Blinnikov}, S.~I. 2016{\natexlab{b}}, Astronomical and Astrophysical
  Transactions, 29, 129

\bibitem[{{Bochenek} {et~al.}(2018){Bochenek}, {Dwarkadas}, {Silverman}, {Fox},
  {Chevalier}, {Smith}, \& {Filippenko}}]{Bochenek18}
{Bochenek}, C.~D., {Dwarkadas}, V.~V., {Silverman}, J.~M., {et~al.} 2018,
  \mnras, 473, 336

\bibitem[{{Boyle} {et~al.}(2008){Boyle}, {Gahbauer}, {H{\"o}ppner},
  {H{\"o}randel}, {Ichimura}, {M{\"u}ller}, {Wolf}, \& {TRACER
  project}}]{BoyleEtal2008}
{Boyle}, P.~J., {Gahbauer}, F., {H{\"o}ppner}, C., {et~al.} 2008, Advances in
  Space Research, 42, 409

\bibitem[{{Branch} \& {Wheeler}(2017)}]{BranchWheeler17}
{Branch}, D. \& {Wheeler}, J.~C. 2017, {Supernova Explosions}

\bibitem[{{Brose} {et~al.}(2016){Brose}, {Telezhinsky}, \& {Pohl}}]{BTP2016}
{Brose}, R., {Telezhinsky}, I., \& {Pohl}, M. 2016, \aap, 593, A20

\bibitem[{{Budnik} {et~al.}(2008){Budnik}, {Katz}, {MacFadyen}, \&
  {Waxman}}]{Budnik2008}
{Budnik}, R., {Katz}, B., {MacFadyen}, A., \& {Waxman}, E. 2008, \apj, 673, 928

\bibitem[{{Buitink} {et~al.}(2016){Buitink}, {Corstanje}, {Falcke},
  {H{\"o}randel}, {Huege}, {Nelles}, {Rachen}, {Rossetto}, {Schellart},
  {Scholten}, {Ter Veen}, {Thoudam}, {Trinh}, {Anderson}, {Asgekar}, {Avruch},
  {Bell}, {Bentum}, {Bernardi}, {Best}, {Bonafede}, {Breitling}, {Broderick},
  {Brouw}, {Br{\"u}ggen}, {Butcher}, {Carbone}, {Ciardi}, {Conway}, {de
  Gasperin}, {de Geus}, {Deller}, {Dettmar}, {van Diepen}, {Duscha},
  {Eisl{\"o}ffel}, {Engels}, {Enriquez}, {Fallows}, {Fender}, {Ferrari},
  {Frieswijk}, {Garrett}, {Grie{\ss}meier}, {Gunst}, {van Haarlem}, {Hassall},
  {Heald}, {Hessels}, {Hoeft}, {Horneffer}, {Iacobelli}, {Intema}, {Juette},
  {Karastergiou}, {Kondratiev}, {Kramer}, {Kuniyoshi}, {Kuper}, {van Leeuwen},
  {Loose}, {Maat}, {Mann}, {Markoff}, {McFadden}, {McKay-Bukowski}, {McKean},
  {Mevius}, {Mulcahy}, {Munk}, {Norden}, {Orru}, {Paas}, {Pandey-Pommier},
  {Pandey}, {Pietka}, {Pizzo}, {Polatidis}, {Reich}, {R{\"o}ttgering},
  {Scaife}, {Schwarz}, {Serylak}, {Sluman}, {Smirnov}, {Stappers}, {Steinmetz},
  {Stewart}, {Swinbank}, {Tagger}, {Tang}, {Tasse}, {Toribio}, {Vermeulen},
  {Vocks}, {Vogt}, {van Weeren}, {Wijers}, {Wijnholds}, {Wise}, {Wucknitz},
  {Yatawatta}, {Zarka}, \& {Zensus}}]{BuitinkEtal2016}
{Buitink}, S., {Corstanje}, A., {Falcke}, H., {et~al.} 2016, \nat, 531, 70

\bibitem[{{Bykov}(2001)}]{Bykov2001}
{Bykov}, A.~M. 2001, Space Science Reviews, 99, 317

\bibitem[{{Bykov}(2014)}]{Bykov2014}
{Bykov}, A.~M. 2014, \aapr, 22, 77

\bibitem[{{Bykov} {et~al.}(2013){Bykov}, {Brandenburg}, {Malkov}, \&
  {Osipov}}]{bbmo13}
{Bykov}, A.~M., {Brandenburg}, A., {Malkov}, M.~A., \& {Osipov}, S.~M. 2013,
  \ssr, 178, 201

\bibitem[{{Bykov} {et~al.}(2015){Bykov}, {Ellison}, {Gladilin}, \&
  {Osipov}}]{BEGO2015MNRAS}
{Bykov}, A.~M., {Ellison}, D.~C., {Gladilin}, P.~E., \& {Osipov}, S.~M. 2015,
  \mnras, 453, 113

\bibitem[{{Bykov} {et~al.}(2017{\natexlab{a}}){Bykov}, {Ellison}, {Gladilin},
  \& {Osipov}}]{Bykov_Clusters2017}
{Bykov}, A.~M., {Ellison}, D.~C., {Gladilin}, P.~E., \& {Osipov}, S.~M.
  2017{\natexlab{a}}, Adv. Space Res. (2017), ArXiv e-prints 1706.01135

\bibitem[{{Bykov} {et~al.}(2017{\natexlab{b}}){Bykov}, {Ellison}, \&
  {Osipov}}]{Bykov_SuperD2017}
{Bykov}, A.~M., {Ellison}, D.~C., \& {Osipov}, S.~M. 2017{\natexlab{b}}, \pre,
  95, 033207

\bibitem[{{Bykov} {et~al.}(2011){Bykov}, {Ellison}, {Osipov}, {Pavlov}, \&
  {Uvarov}}]{beopu2011}
{Bykov}, A.~M., {Ellison}, D.~C., {Osipov}, S.~M., {Pavlov}, G.~G., \&
  {Uvarov}, Y.~A. 2011, \apjl, 735, L40

\bibitem[{{Bykov} {et~al.}(2014){Bykov}, {Ellison}, {Osipov}, \&
  {Vladimirov}}]{Bykov3inst2014}
{Bykov}, A.~M., {Ellison}, D.~C., {Osipov}, S.~M., \& {Vladimirov}, A.~E. 2014,
  ApJ, 789, 137

\bibitem[{{Bykov} {et~al.}(2017{\natexlab{c}}){Bykov}, {Osipov}, \&
  {Ellison}}]{BykovTransRelMFP}
{Bykov}, A.~M., {Osipov}, S.~M., \& {Ellison}, D.~C. 2017{\natexlab{c}}, In
  preparation

\bibitem[{{Bykov} \& {Toptygin}(1990)}]{BT1990}
{Bykov}, A.~M. \& {Toptygin}, I.~N. 1990, Sov. Phys. JETP, 71

\bibitem[{{Caprioli} {et~al.}(2009){Caprioli}, {Blasi}, \& {Amato}}]{CBA2009}
{Caprioli}, D., {Blasi}, P., \& {Amato}, E. 2009, \mnras, 396, 2065

\bibitem[{{Cassam-Chena{\"{\i}}} {et~al.}(2008){Cassam-Chena{\"{\i}}},
  {Hughes}, {Reynoso}, {Badenes}, \& {Moffett}}]{Cassam2008}
{Cassam-Chena{\"{\i}}}, G., {Hughes}, J.~P., {Reynoso}, E.~M., {Badenes}, C.,
  \& {Moffett}, D. 2008, \apj, 680, 1180

\bibitem[{{Castro} {et~al.}(2012){Castro}, {Slane}, {Ellison}, \&
  {Patnaude}}]{CSEP2012}
{Castro}, D., {Slane}, P., {Ellison}, D.~C., \& {Patnaude}, D.~J. 2012, \apj,
  756, 88

\bibitem[{{Cesarsky} \& {Montmerle}(1983)}]{cm83}
{Cesarsky}, C.~J. \& {Montmerle}, T. 1983, \ssr, 36, 173

\bibitem[{{Chakraborti} \& {Ray}(2011)}]{Chakraborti11}
{Chakraborti}, S. \& {Ray}, A. 2011, \apj, 729, 57

\bibitem[{{Chakraborti} {et~al.}(2011){Chakraborti}, {Ray}, {Soderberg},
  {Loeb}, \& {Chandra}}]{Chakraborti2011}
{Chakraborti}, S., {Ray}, A., {Soderberg}, A.~M., {Loeb}, A., \& {Chandra}, P.
  2011, Nature Communications, 2

\bibitem[{{Chandra} {et~al.}(2015){Chandra}, {Chevalier}, {Chugai}, {Fransson},
  \& {Soderberg}}]{Chandra15}
{Chandra}, P., {Chevalier}, R.~A., {Chugai}, N., {Fransson}, C., \&
  {Soderberg}, A.~M. 2015, \apj, 810, 32

\bibitem[{{Chevalier}(1982)}]{ch82}
{Chevalier}, R.~A. 1982, \apj, 258, 790

\bibitem[{{Chevalier}(2012)}]{Chevalier2012}
{Chevalier}, R.~A. 2012, \apjl, 752, L2

\bibitem[{{Chevalier} \& {Clegg}(1985)}]{chev_clegg85}
{Chevalier}, R.~A. \& {Clegg}, A.~W. 1985, \nat, 317, 44

\bibitem[{{Chevalier} \& {Irwin}(2011)}]{ChevalierI2011}
{Chevalier}, R.~A. \& {Irwin}, C.~M. 2011, \apjl, 729, L6

\bibitem[{{Chugai} \& {Danziger}(1994)}]{Chugai1994}
{Chugai}, N.~N. \& {Danziger}, I.~J. 1994, \mnras, 268, 173

\bibitem[{{Clark} {et~al.}(2008){Clark}, {Muno}, {Negueruela}, {Dougherty},
  {Crowther}, {Goodwin}, \& {de Grijs}}]{Clark2008}
{Clark}, J.~S., {Muno}, M.~P., {Negueruela}, I., {et~al.} 2008, \aap, 477, 147

\bibitem[{{Comer{\'o}n} {et~al.}(2016){Comer{\'o}n}, {Djupvik}, {Schneider}, \&
  {Pasquali}}]{ComeronEtal2016}
{Comer{\'o}n}, F., {Djupvik}, A.~A., {Schneider}, N., \& {Pasquali}, A. 2016,
  \aap, 586, A46

\bibitem[{{Cox}(2005)}]{Cox2005}
{Cox}, D.~P. 2005, \araa, 43, 337

\bibitem[{{De Angelis} {et~al.}(2017){De Angelis}, {Tatischeff}, {Tavani},
  {Oberlack}, {Grenier}, {Hanlon}, {Walter}, {Argan}, {von Ballmoos},
  {Bulgarelli}, {Donnarumma}, {Hernanz}, {Kuvvetli}, {Pearce}, {Zdziarski},
  {Aboudan}, {Ajello}, {Ambrosi}, {Bernard}, {Bernardini}, {Bonvicini},
  {Brogna}, {Branchesi}, {Budtz-Jorgensen}, {Bykov}, {Campana}, {Cardillo},
  {Coppi}, {De Martino}, {Diehl}, {Doro}, {Fioretti}, {Funk}, {Ghisellini},
  {Grove}, {Hamadache}, {Hartmann}, {Hayashida}, {Isern}, {Kanbach}, {Kiener},
  {Kn{\"o}dlseder}, {Labanti}, {Laurent}, {Limousin}, {Longo}, {Mannheim},
  {Marisaldi}, {Martinez}, {Mazziotta}, {McEnery}, {Mereghetti}, {Minervini},
  {Moiseev}, {Morselli}, {Nakazawa}, {Orleanski}, {Paredes}, {Patricelli},
  {Peyr{\'e}}, {Piano}, {Pohl}, {Ramarijaona}, {Rando}, {Reichardt},
  {Roncadelli}, {Silva}, {Tavecchio}, {Thompson}, {Turolla}, {Ulyanov},
  {Vacchi}, {Wu}, \& {Zoglauer}}]{DeAngelis2017}
{De Angelis}, A., {Tatischeff}, V., {Tavani}, M., {et~al.} 2017, Experimental
  Astronomy, 44, 25

\bibitem[{Decourchelle {et~al.}(2000)Decourchelle, Ellison, \&
  Ballet}]{DEB2000}
Decourchelle, A., Ellison, D.~C., \& Ballet, J. 2000, ApJ, 543, L57

\bibitem[{{Di Sciascio} \& {LHAASO Collaboration}(2016)}]{Disciascio16}
{Di Sciascio}, G. \& {LHAASO Collaboration}. 2016, Nuclear and Particle Physics
  Proceedings, 279, 166

\bibitem[{{Drury}(2011)}]{Drury2011}
{Drury}, L.~O. 2011, \mnras, 415, 1807

\bibitem[{{Dwarkadas}(2005)}]{Dwarkadas2005}
{Dwarkadas}, V.~V. 2005, \apj, 630, 892

\bibitem[{{Dwarkadas} \& {Chevalier}(1998)}]{DC98}
{Dwarkadas}, V.~V. \& {Chevalier}, R.~A. 1998, \apj, 497, 807

\bibitem[{{Ellison} {et~al.}(1995){Ellison}, {Baring}, \& {Jones}}]{EBJ95}
{Ellison}, D.~C., {Baring}, M.~G., \& {Jones}, F.~C. 1995, \apj, 453, 873

\bibitem[{{Ellison} \& {Bykov}(2011)}]{EB2011}
{Ellison}, D.~C. \& {Bykov}, A.~M. 2011, ApJ, 731, 87

\bibitem[{{Ellison} {et~al.}(1997){Ellison}, {Drury}, \& {Meyer}}]{EDM97}
{Ellison}, D.~C., {Drury}, L.~O., \& {Meyer}, J. 1997, \apj, 487, 197

\bibitem[{{Ellison} {et~al.}(1990){Ellison}, {Moebius}, \& {Paschmann}}]{EMP90}
{Ellison}, D.~C., {Moebius}, E., \& {Paschmann}, G. 1990, \apj, 352, 376

\bibitem[{{Ellison} {et~al.}(2007){Ellison}, {Patnaude}, {Slane}, {Blasi}, \&
  {Gabici}}]{EPSBG2007}
{Ellison}, D.~C., {Patnaude}, D.~J., {Slane}, P., {Blasi}, P., \& {Gabici}, S.
  2007, \apj, 661, 879

\bibitem[{{Ellison} \& {Ramaty}(1985)}]{ER85}
{Ellison}, D.~C. \& {Ramaty}, R. 1985, \apj, 298, 400

\bibitem[{{Ellison} {et~al.}(2012){Ellison}, {Slane}, {Patnaude}, \&
  {Bykov}}]{ESPB2012}
{Ellison}, D.~C., {Slane}, P., {Patnaude}, D.~J., \& {Bykov}, A.~M. 2012, ApJ,
  744, 39

\bibitem[{{Ellison} {et~al.}(2013){Ellison}, {Warren}, \& {Bykov}}]{EWB2013}
{Ellison}, D.~C., {Warren}, D.~C., \& {Bykov}, A.~M. 2013, ApJ, 776, 46

\bibitem[{{Eriksen} {et~al.}(2011){Eriksen}, {Hughes}, {Badenes}, {Fesen},
  {Ghavamian}, {Moffett}, {Plucinksy}, {Rakowski}, {Reynoso}, \&
  {Slane}}]{EriksenEtal2011}
{Eriksen}, K.~A., {Hughes}, J.~P., {Badenes}, C., {et~al.} 2011, \apjl, 728,
  L28+

\bibitem[{{Falk} \& {Arnett}(1977)}]{FalkArnett1977}
{Falk}, S.~W. \& {Arnett}, W.~D. 1977, \aaps, 33, 515

\bibitem[{{Farber} {et~al.}(2017){Farber}, {Ruszkowski}, {Yang}, \&
  {Zweibel}}]{FRYZ17}
{Farber}, R., {Ruszkowski}, M., {Yang}, H.-Y.~K., \& {Zweibel}, E.~G. 2017,
  ArXiv e-prints, 1707.04579

\bibitem[{{Fermi}(1949)}]{Fermi_PR49}
{Fermi}, E. 1949, Physical Review, 75, 1169

\bibitem[{{Fermi}(1954)}]{Fermi_ApJ54}
{Fermi}, E. 1954, \apj, 119, 1

\bibitem[{{Ferrand} {et~al.}(2012){Ferrand}, {Decourchelle}, \&
  {Safi-Harb}}]{FDS2012}
{Ferrand}, G., {Decourchelle}, A., \& {Safi-Harb}, S. 2012, \apj, 760, 34

\bibitem[{{Ferrand} \& {Marcowith}(2010)}]{Ferrand2010}
{Ferrand}, G. \& {Marcowith}, A. 2010, \aap, 510, A101

\bibitem[{{Fiorito} {et~al.}(1990){Fiorito}, {Eichler}, \&
  {Ellison}}]{Fiorito1990}
{Fiorito}, R.~B., {Eichler}, D., \& {Ellison}, D.~C. 1990, \apj, 364, 582

\bibitem[{{Fransson} \& {Bj{\"o}rnsson}(1998)}]{Fransson98}
{Fransson}, C. \& {Bj{\"o}rnsson}, C.-I. 1998, \apj, 509, 861

\bibitem[{{Fujita} {et~al.}(2017){Fujita}, {Murase}, \& {Kimura}}]{Fujita2017}
{Fujita}, Y., {Murase}, K., \& {Kimura}, S.~S. 2017, \jcap, 4, 037

\bibitem[{{Funk}(2015)}]{Funk2015}
{Funk}, S. 2015, Annual Review of Nuclear and Particle Science, 65, 245

\bibitem[{{Gabici} \& {Aharonian}(2014)}]{Gabici14}
{Gabici}, S. \& {Aharonian}, F.~A. 2014, \mnras, 445, L70

\bibitem[{{Gal-Yam}(2012)}]{GalYam2012}
{Gal-Yam}, A. 2012, Science, 337, 927

\bibitem[{{Ginzburg} \& {Syrovatskii}(1964)}]{Ginzburg1964}
{Ginzburg}, V.~L. \& {Syrovatskii}, S.~I. 1964, {The Origin of Cosmic Rays}

\bibitem[{{Giuliani} {et~al.}(2011){Giuliani}, {Cardillo}, {Tavani}, {Fukui},
  {Yoshiike}, {Torii}, {Dubner}, {Castelletti}, {Barbiellini}, {Bulgarelli},
  {Caraveo}, {Costa}, {Cattaneo}, {Chen}, {Contessi}, {Del Monte},
  {Donnarumma}, {Evangelista}, {Feroci}, {Gianotti}, {Lazzarotto}, {Lucarelli},
  {Longo}, {Marisaldi}, {Mereghetti}, {Pacciani}, {Pellizzoni}, {Piano},
  {Picozza}, {Pittori}, {Pucella}, {Rapisarda}, {Rappoldi}, {Sabatini},
  {Soffitta}, {Striani}, {Trifoglio}, {Trois}, {Vercellone}, {Verrecchia},
  {Vittorini}, {Colafrancesco}, {Giommi}, \& {Bignami}}]{GiulianiEtal2011}
{Giuliani}, A., {Cardillo}, M., {Tavani}, M., {et~al.} 2011, \apjl, 742, L30

\bibitem[{{Grenier} {et~al.}(2015){Grenier}, {Black}, \&
  {Strong}}]{Grenier2015}
{Grenier}, I.~A., {Black}, J.~H., \& {Strong}, A.~W. 2015, \araa, 53, 199

\bibitem[{{Guo} {et~al.}(2017){Guo}, {Tian}, {Wang}, {Li}, \&
  {Chen}}]{GuoEtal2017}
{Guo}, Y.-Q., {Tian}, Z., {Wang}, Z., {Li}, H.-J., \& {Chen}, T.-L. 2017, \apj,
  836, 233

\bibitem[{{Heiles}(1990)}]{heiles90}
{Heiles}, C. 1990, \apj, 354, 483

\bibitem[{{Helder} {et~al.}(2012){Helder}, {Vink}, {Bykov}, {Ohira}, {Raymond},
  \& {Terrier}}]{helder12}
{Helder}, E.~A., {Vink}, J., {Bykov}, A.~M., {et~al.} 2012, \ssr, 173, 369

\bibitem[{{HESS Collaboration} {et~al.}(2016){HESS Collaboration},
  {Abramowski}, {Aharonian}, {Benkhali}, {Akhperjanian}, {Ang{\"u}ner},
  {Backes}, {Balzer}, {Becherini}, {Tjus}, \& et~al.}]{HESS2016}
{HESS Collaboration}, {Abramowski}, A., {Aharonian}, F., {et~al.} 2016, \nat,
  531, 476

\bibitem[{{Hillas}(2005)}]{Hillas2005}
{Hillas}, A.~M. 2005, Journal of Physics G Nuclear Physics, 31, 95

\bibitem[{{Horesh} {et~al.}(2013){Horesh}, {Stockdale}, {Fox}, {Frail},
  {Carpenter}, {Kulkarni}, {Ofek}, {Gal-Yam}, {Kasliwal}, {Arcavi}, {Quimby},
  {Cenko}, {Nugent}, {Bloom}, {Law}, {Poznanski}, {Gorbikov}, {Polishook},
  {Yaron}, {Ryder}, {Weiler}, {Bauer}, {Van Dyk}, {Immler}, {Panagia},
  {Pooley}, \& {Kassim}}]{Horesh13}
{Horesh}, A., {Stockdale}, C., {Fox}, D.~B., {et~al.} 2013, \mnras, 436, 1258

\bibitem[{{Hughes} {et~al.}(2000){Hughes}, {Rakowski}, \&
  {Decourchelle}}]{HRD2000}
{Hughes}, J.~P., {Rakowski}, C.~E., \& {Decourchelle}, A. 2000, \apjl, 543, L61

\bibitem[{{Jones} \& {Ellison}(1991)}]{je91}
{Jones}, F.~C. \& {Ellison}, D.~C. 1991, Space Science Reviews, 58, 259

\bibitem[{{Kang}(2013)}]{Kang2013}
{Kang}, H. 2013, Journal of Korean Astronomical Society, 46, 49

\bibitem[{{Kang} {et~al.}(2009){Kang}, {Ryu}, \& {Jones}}]{KJ2009}
{Kang}, H., {Ryu}, D., \& {Jones}, T.~W. 2009, \apj, 695, 1273

\bibitem[{{Katsuta} {et~al.}(2017){Katsuta}, {Uchiyama}, \&
  {Funk}}]{Katsuta2017}
{Katsuta}, J., {Uchiyama}, Y., \& {Funk}, S. 2017, \apj, 839, 129

\bibitem[{{Katz} {et~al.}(2011){Katz}, {Sapir}, \& {Waxman}}]{Katz2011}
{Katz}, B., {Sapir}, N., \& {Waxman}, E. 2011, ArXiv e-prints

\bibitem[{{Kimani} {et~al.}(2016){Kimani}, {Sendlinger}, {Brunthaler},
  {Menten}, {Mart{\'{\i}}-Vidal}, {Henkel}, {Falcke}, {Muxlow}, {Beswick}, \&
  {Bower}}]{Kimani16}
{Kimani}, N., {Sendlinger}, K., {Brunthaler}, A., {et~al.} 2016, \aap, 593, A18

\bibitem[{{Kn{\"o}dlseder}(2000)}]{Knodlseder2000}
{Kn{\"o}dlseder}, J. 2000, \aap, 360, 539

\bibitem[{{Krauss} {et~al.}(2012){Krauss}, {Soderberg}, {Chomiuk}, {Zauderer},
  {Brunthaler}, {Bietenholz}, {Chevalier}, {Fransson}, \& {Rupen}}]{Krauss12}
{Krauss}, M.~I., {Soderberg}, A.~M., {Chomiuk}, L., {et~al.} 2012, \apjl, 750,
  L40

\bibitem[{{Krumholz}(2017)}]{Krumholz2017}
{Krumholz}, M. 2017, {Star Formation}

\bibitem[{{Krymskii}(1977)}]{Kry77}
{Krymskii}, G.~F. 1977, Akademiia Nauk SSSR Doklady, 234, 1306

\bibitem[{{Kulkarni} {et~al.}(1998){Kulkarni}, {Frail}, {Wieringa}, {Ekers},
  {Sadler}, {Wark}, {Higdon}, {Phinney}, \& {Bloom}}]{Kulkarni1998}
{Kulkarni}, S.~R., {Frail}, D.~A., {Wieringa}, M.~H., {et~al.} 1998, \nat, 395,
  663

\bibitem[{{Kundu} {et~al.}(2017){Kundu}, {Lundqvist}, {P{\'e}rez-Torres},
  {Herrero-Illana}, \& {Alberdi}}]{KunduEtal2017}
{Kundu}, E., {Lundqvist}, P., {P{\'e}rez-Torres}, M.~A., {Herrero-Illana}, R.,
  \& {Alberdi}, A. 2017, \apj, 842, 17

\bibitem[{{Lada} \& {Lada}(2003)}]{Lada2003}
{Lada}, C.~J. \& {Lada}, E.~A. 2003, \araa, 41, 57

\bibitem[{{Lagage} \& {Cesarsky}(1983)}]{LC83}
{Lagage}, P.~O. \& {Cesarsky}, C.~J. 1983, \aap, 125, 249

\bibitem[{{Lee} {et~al.}(2012){Lee}, {Ellison}, \& {Nagataki}}]{LEN2012}
{Lee}, S.-H., {Ellison}, D.~C., \& {Nagataki}, S. 2012, \apj, 750, 156

\bibitem[{{Lemoine}(2013)}]{Lemoine2013}
{Lemoine}, M. 2013, in Journal of Physics Conference Series, Vol. 409, Journal
  of Physics Conference Series, 012007

\bibitem[{{Li} {et~al.}(2011){Li}, {Leaman}, {Chornock}, {Filippenko},
  {Poznanski}, {Ganeshalingam}, {Wang}, {Modjaz}, {Jha}, {Foley}, \&
  {Smith}}]{LiEtal2011}
{Li}, W., {Leaman}, J., {Chornock}, R., {et~al.} 2011, \mnras, 412, 1441

\bibitem[{{Lingenfelter}(2017)}]{Lingenfelter17}
{Lingenfelter}, R.~E. 2017, Advances in Space Research,
  doi:10.1016/j.asr.2017.04.006

\bibitem[{{Lipari}(2017)}]{Lipari2017}
{Lipari}, P. 2017, ArXiv e-prints 1707.02504

\bibitem[{{Lithwick} \& {Goldreich}(2001)}]{Lithwick2001}
{Lithwick}, Y. \& {Goldreich}, P. 2001, \apj, 562, 279

\bibitem[{{Liu} {et~al.}(2016){Liu}, {Wang}, {Prosekin}, \&
  {Chang}}]{LiuEtal2016}
{Liu}, R.-Y., {Wang}, X.-Y., {Prosekin}, A., \& {Chang}, X.-C. 2016, \apj, 833,
  200

\bibitem[{{Loeb} \& {Waxman}(2006)}]{LW06}
{Loeb}, A. \& {Waxman}, E. 2006, \jcap, 5, 003

\bibitem[{{Mac Low} \& {McCray}(1988)}]{maclow_mccray88}
{Mac Low}, M.-M. \& {McCray}, R. 1988, \apj, 324, 776

\bibitem[{{Malkov}(2017)}]{Malkov2017}
{Malkov}, M. 2017, ArXiv e-prints

\bibitem[{{Malkov} {et~al.}(2013){Malkov}, {Diamond}, {Sagdeev}, {Aharonian},
  \& {Moskalenko}}]{MalkovEtal2013}
{Malkov}, M.~A., {Diamond}, P.~H., {Sagdeev}, R.~Z., {Aharonian}, F.~A., \&
  {Moskalenko}, I.~V. 2013, \apj, 768, 73

\bibitem[{{Malkov} \& {Drury}(2001)}]{MD2001}
{Malkov}, M.~A. \& {Drury}, L. 2001, Reports of Progress in Physics, 64, 429

\bibitem[{{Marcaide} {et~al.}(2009){Marcaide}, {Mart{\'{\i}}-Vidal},
  {Perez-Torres}, {Alberdi}, {Guirado}, {Ros}, \& {Weiler}}]{Marcaide09}
{Marcaide}, J.~M., {Mart{\'{\i}}-Vidal}, I., {Perez-Torres}, M.~A., {et~al.}
  2009, \aap, 503, 869

\bibitem[{{Marcowith} {et~al.}(2016){Marcowith}, {Bret}, {Bykov}, {Dieckman},
  {O'C Drury}, {Lemb{\`e}ge}, {Lemoine}, {Morlino}, {Murphy}, {Pelletier},
  {Plotnikov}, {Reville}, {Riquelme}, {Sironi}, \& {Stockem
  Novo}}]{MarcowithEtal2016}
{Marcowith}, A., {Bret}, A., {Bykov}, A., {et~al.} 2016, Reports on Progress in
  Physics, 79, 046901

\bibitem[{{Marcowith} \& {Casse}(2010)}]{Marcowith10}
{Marcowith}, A. \& {Casse}, F. 2010, \aap, 515, A90

\bibitem[{{Marcowith} {et~al.}(2014){Marcowith}, {Renaud}, {Dwarkadas}, \&
  {Tatischeff}}]{Marcowith14}
{Marcowith}, A., {Renaud}, M., {Dwarkadas}, V., \& {Tatischeff}, V. 2014,
  Nuclear Physics B Proceedings Supplements, 256, 94

\bibitem[{{Margutti} {et~al.}(2014){Margutti}, {Milisavljevic}, {Soderberg},
  {Guidorzi}, {Morsony}, {Sanders}, {Chakraborti}, {Ray}, {Kamble}, {Drout},
  {Parrent}, {Zauderer}, \& {Chomiuk}}]{Margutti2014}
{Margutti}, R., {Milisavljevic}, D., {Soderberg}, A.~M., {et~al.} 2014, \apj,
  797, 107

\bibitem[{{Mart{\'{\i}}-Vidal} {et~al.}(2011){Mart{\'{\i}}-Vidal}, {Marcaide},
  {Alberdi}, {Guirado}, {P{\'e}rez-Torres}, \& {Ros}}]{Marti11}
{Mart{\'{\i}}-Vidal}, I., {Marcaide}, J.~M., {Alberdi}, A., {et~al.} 2011,
  \aap, 526, A143

\bibitem[{{Martin} {et~al.}(2010){Martin}, {Kn{\"o}dlseder}, {Meynet}, \&
  {Diehl}}]{Martinea10}
{Martin}, P., {Kn{\"o}dlseder}, J., {Meynet}, G., \& {Diehl}, R. 2010, \aap,
  511, A86

\bibitem[{{Meyer} {et~al.}(1997){Meyer}, {Drury}, \& {Ellison}}]{MDE97}
{Meyer}, J., {Drury}, L.~O., \& {Ellison}, D.~C. 1997, \apj, 487, 182

\bibitem[{{Meyer} \& {Ellison}(1999)}]{ME1999}
{Meyer}, J.~P. \& {Ellison}, D.~C. 1999, ArXiv Astrophysics e-prints

\bibitem[{{Moriya} {et~al.}(2013){Moriya}, {Blinnikov}, {Tominaga}, {Yoshida},
  {Tanaka}, {Maeda}, \& {Nomoto}}]{MoriyaEtal2013}
{Moriya}, T.~J., {Blinnikov}, S.~I., {Tominaga}, N., {et~al.} 2013, \mnras,
  428, 1020

\bibitem[{{Moriya} \& {Maeda}(2014)}]{Moriya2014}
{Moriya}, T.~J. \& {Maeda}, K. 2014, \apjl, 790, L16

\bibitem[{{Morozova} {et~al.}(2017){Morozova}, {Piro}, \&
  {Valenti}}]{Morozova2017}
{Morozova}, V., {Piro}, A.~L., \& {Valenti}, S. 2017, \apj, 838, 28

\bibitem[{{Murase} {et~al.}(2011){Murase}, {Thompson}, {Lacki}, \&
  {Beacom}}]{MuraseEtal2011}
{Murase}, K., {Thompson}, T.~A., {Lacki}, B.~C., \& {Beacom}, J.~F. 2011, \prd,
  84, 043003

\bibitem[{{Murase} {et~al.}(2014){Murase}, {Thompson}, \& {Ofek}}]{Murase2014}
{Murase}, K., {Thompson}, T.~A., \& {Ofek}, E.~O. 2014, \mnras, 440, 2528

\bibitem[{{Nugis} \& {Lamers}(2000)}]{Nugis_Lamers2000}
{Nugis}, T. \& {Lamers}, H.~J.~G.~L.~M. 2000, \aap, 360, 227

\bibitem[{Ohm(2012)}]{Ohm2012}
Ohm, S. 2012, AIP Conf. Proc., 1505, 64

\bibitem[{{Ohm}(2016)}]{ohm16}
{Ohm}, S. 2016, Comptes Rendus Physique, 17, 585

\bibitem[{{Parizot} {et~al.}(2006){Parizot}, {Marcowith}, {Ballet}, \&
  {Gallant}}]{Parizot2006}
{Parizot}, E., {Marcowith}, A., {Ballet}, J., \& {Gallant}, Y.~A. 2006, \aap,
  453, 387

\bibitem[{{Parizot} {et~al.}(2004){Parizot}, {Marcowith}, {van der Swaluw},
  {Bykov}, \& {Tatischeff}}]{parizot04}
{Parizot}, E., {Marcowith}, A., {van der Swaluw}, E., {Bykov}, A.~M., \&
  {Tatischeff}, V. 2004, \aap, 424, 747

\bibitem[{{Patnaude} {et~al.}(2017){Patnaude}, {Lee}, {Slane}, {Badenes},
  {Nagataki}, {Ellison}, \& {Milisavljevic}}]{PatnaudeEtal2017}
{Patnaude}, D.~J., {Lee}, S.-H., {Slane}, P.~O., {et~al.} 2017, ArXiv e-prints

\bibitem[{{Perez-Torres} {et~al.}(2015){Perez-Torres}, {Alberdi}, {Beswick},
  {Lundqvist}, {Herrero-Illana}, {Romero-Ca{\~n}izales}, {Ryder}, {della
  Valle}, {Conway}, {Marcaide}, {Mattila}, {Murphy}, \& {Ros}}]{Perez15}
{Perez-Torres}, M., {Alberdi}, A., {Beswick}, R.~J., {et~al.} 2015, Advancing
  Astrophysics with the Square Kilometre Array (AASKA14), 60

\bibitem[{{Petropoulou} {et~al.}(2017){Petropoulou}, {Coenders},
  {Vasilopoulos}, {Kamble}, \& {Sironi}}]{PetropoulouEtal2017}
{Petropoulou}, M., {Coenders}, S., {Vasilopoulos}, G., {Kamble}, A., \&
  {Sironi}, L. 2017, \mnras, 470, 1881

\bibitem[{{Portegies Zwart} {et~al.}(2010){Portegies Zwart}, {McMillan}, \&
  {Gieles}}]{Portegies_Zwart2010}
{Portegies Zwart}, S.~F., {McMillan}, S.~L.~W., \& {Gieles}, M. 2010, \araa,
  48, 431

\bibitem[{{Ptuskin}(2012)}]{Ptuskin2012}
{Ptuskin}, V. 2012, Astroparticle Physics, 39, 44

\bibitem[{{Ptuskin} \& {Zirakashvili}(2005)}]{PZ2005a}
{Ptuskin}, V.~S. \& {Zirakashvili}, V.~N. 2005, \aap, 429, 755

\bibitem[{{Raymond}(2017)}]{Raymond2017}
{Raymond}, J.~C. 2017, Space Science Reviews (in press)

\bibitem[{{Recchia} {et~al.}(2017){Recchia}, {Blasi}, \& {Morlino}}]{RBM17}
{Recchia}, S., {Blasi}, P., \& {Morlino}, G. 2017, \mnras, 470, 865

\bibitem[{{Reynolds}(2008)}]{Reynolds08}
{Reynolds}, S.~P. 2008, \araa, 46, 89

\bibitem[{{Ross} \& {Dwarkadas}(2017)}]{Ross17}
{Ross}, M. \& {Dwarkadas}, V.~V. 2017, \aj, 153, 246

\bibitem[{{Rothenflug} {et~al.}(2004){Rothenflug}, {Ballet}, {Dubner},
  {Giacani}, {Decourchelle}, \& {Ferrando}}]{RothenflugEtal2004}
{Rothenflug}, R., {Ballet}, J., {Dubner}, G., {et~al.} 2004, \aap, 425, 121

\bibitem[{{Rygl} {et~al.}(2012){Rygl}, {Brunthaler}, {Sanna}, {Menten}, {Reid},
  {van Langevelde}, {Honma}, {Torstensson}, \& {Fujisawa}}]{RyglEtal2012}
{Rygl}, K.~L.~J., {Brunthaler}, A., {Sanna}, A., {et~al.} 2012, \aap, 539, A79

\bibitem[{{Schure} \& {Bell}(2013)}]{SB13}
{Schure}, K.~M. \& {Bell}, A.~R. 2013, \mnras, 435, 1174

\bibitem[{{Schure} {et~al.}(2012){Schure}, {Bell}, {O'C Drury}, \&
  {Bykov}}]{SchureEtal2012}
{Schure}, K.~M., {Bell}, A.~R., {O'C Drury}, L., \& {Bykov}, A.~M. 2012, \ssr,
  173, 491

\bibitem[{{Simoni} {et~al.}(2017){Simoni}, {Maxted}, {Renaud}, \&
  {Vink}}]{Simoni17}
{Simoni}, R., {Maxted}, N., {Renaud}, M., \& {Vink}, J. 2017, in IAU Symposium,
  Vol. 331, Supernova 1987A:30 years later - Cosmic Rays and Nuclei from
  Supernovae and their Aftermaths, ed. A.~{Marcowith}, M.~{Renaud},
  G.~{Dubner}, A.~{Ray}, \& A.~{Bykov}, 325--328

\bibitem[{{Slane} {et~al.}(2015{\natexlab{a}}){Slane}, {Bykov}, {Ellison},
  {Dubner}, \& {Castro}}]{SlaneSSRv15}
{Slane}, P., {Bykov}, A., {Ellison}, D.~C., {Dubner}, G., \& {Castro}, D.
  2015{\natexlab{a}}, \ssr, 188, 187

\bibitem[{{Slane} {et~al.}(2015{\natexlab{b}}){Slane}, {Lee}, {Ellison},
  {Patnaude}, {Hughes}, {Eriksen}, {Castro}, \& {Nagataki}}]{Slane2015Err}
{Slane}, P., {Lee}, S.-H., {Ellison}, D.~C., {et~al.} 2015{\natexlab{b}}, \apj,
  799, 238

\bibitem[{{Slysh}(1992)}]{Slysh92}
{Slysh}, V.~I. 1992, Astronomical and Astrophysical Transactions, 1, 171

\bibitem[{{Smith}(2014)}]{Smith14}
{Smith}, N. 2014, \araa, 52, 487

\bibitem[{{Smith}(2017)}]{Smith2017}
{Smith}, N. 2017, Philosophical Transactions of the Royal Society of London
  Series A, 375, 20160268

\bibitem[{{Soderberg} {et~al.}(2010){Soderberg}, {Chakraborti}, {Pignata},
  {Chevalier}, {Chandra}, {Ray}, {Wieringa}, {Copete}, {Chaplin},
  {Connaughton}, {Barthelmy}, {Bietenholz}, {Chugai}, {Stritzinger}, {Hamuy},
  {Fransson}, {Fox}, {Levesque}, {Grindlay}, {Challis}, {Foley}, {Kirshner},
  {Milne}, \& {Torres}}]{Soderberg2010}
{Soderberg}, A.~M., {Chakraborti}, S., {Pignata}, G., {et~al.} 2010, \nat, 463,
  513

\bibitem[{{Soderberg} {et~al.}(2005){Soderberg}, {Kulkarni}, {Berger},
  {Chevalier}, {Frail}, {Fox}, \& {Walker}}]{Soderberg05}
{Soderberg}, A.~M., {Kulkarni}, S.~R., {Berger}, E., {et~al.} 2005, \apj, 621,
  908

\bibitem[{{Soderberg} {et~al.}(2006){Soderberg}, {Kulkarni}, {Nakar}, {Berger},
  {Cameron}, {Fox}, {Frail}, {Gal-Yam}, {Sari}, {Cenko}, {Kasliwal},
  {Chevalier}, {Piran}, {Price}, {Schmidt}, {Pooley}, {Moon}, {Penprase},
  {Ofek}, {Rau}, {Gehrels}, {Nousek}, {Burrows}, {Persson}, \&
  {McCarthy}}]{Soderberg2006}
{Soderberg}, A.~M., {Kulkarni}, S.~R., {Nakar}, E., {et~al.} 2006, \nat, 442,
  1014

\bibitem[{{Sorokina} {et~al.}(2016){Sorokina}, {Blinnikov}, {Nomoto}, {Quimby},
  \& {Tolstov}}]{SorokinaEtal2016}
{Sorokina}, E., {Blinnikov}, S., {Nomoto}, K., {Quimby}, R., \& {Tolstov}, A.
  2016, \apj, 829, 17

\bibitem[{{Sveshnikova}(2003)}]{Sveshnikova2003}
{Sveshnikova}, L.~G. 2003, \aap, 409, 799

\bibitem[{{Tatischeff}(2009)}]{Tatischeff09}
{Tatischeff}, V. 2009, \aap, 499, 191

\bibitem[{{Tavani} {et~al.}(2010){Tavani}, {Giuliani}, {Chen}, {Argan},
  {Barbiellini}, {Bulgarelli}, {Caraveo}, {Cattaneo}, {Cocco}, {Contessi},
  {D'Ammando}, {Costa}, {De Paris}, {Del Monte}, {Di Cocco}, {Donnarumma},
  {Evangelista}, {Ferrari}, {Feroci}, {Fuschino}, {Galli}, {Gianotti},
  {Labanti}, {Lapshov}, {Lazzarotto}, {Lipari}, {Longo}, {Marisaldi},
  {Mastropietro}, {Mereghetti}, {Morelli}, {Moretti}, {Morselli}, {Pacciani},
  {Pellizzoni}, {Perotti}, {Piano}, {Picozza}, {Pilia}, {Pucella}, {Prest},
  {Rapisarda}, {Rappoldi}, {Scalise}, {Rubini}, {Sabatini}, {Striani},
  {Soffitta}, {Trifoglio}, {Trois}, {Vallazza}, {Vercellone}, {Vittorini},
  {Zambra}, {Zanello}, {Pittori}, {Verrecchia}, {Santolamazza}, {Giommi},
  {Colafrancesco}, {Antonelli}, \& {Salotti}}]{Tavani2010}
{Tavani}, M., {Giuliani}, A., {Chen}, A.~W., {et~al.} 2010, \apjl, 710, L151

\bibitem[{{Telezhinsky} {et~al.}(2011){Telezhinsky}, {Dwarkadas}, \&
  {Pohl}}]{TDP2011}
{Telezhinsky}, I., {Dwarkadas}, V.~V., \& {Pohl}, M. 2011, ArXiv e-prints

\bibitem[{{Thoudam et. al.}(2016)}]{Thoudam2016}
{Thoudam et. al.}, S. 2016, \aap, 595, A33

\bibitem[{{Uchiyama} {et~al.}(2010){Uchiyama}, {Blandford}, {Funk}, {Tajima},
  \& {Tanaka}}]{uchiyamaea10}
{Uchiyama}, Y., {Blandford}, R.~D., {Funk}, S., {Tajima}, H., \& {Tanaka}, T.
  2010, \apjl, 723, L122

\bibitem[{{ud-Doula} \& {Owocki}(2002)}]{ud-Doula2002}
{ud-Doula}, A. \& {Owocki}, S.~P. 2002, \apj, 576, 413

\bibitem[{{van Marle} {et~al.}(2017){van Marle}, {Casse}, \&
  {Marcowith}}]{Marle_Casse_Marcowith2017}
{van Marle}, A.~J., {Casse}, F., \& {Marcowith}, A. 2017, ArXiv e-prints

\bibitem[{{Vink}(2012)}]{vink12}
{Vink}, J. 2012, \aapr, 20, 49

\bibitem[{{Vink}(2017)}]{Vink2017}
{Vink}, J.~S. 2017, ArXiv e-prints

\bibitem[{{Walder} {et~al.}(2012){Walder}, {Folini}, \&
  {Meynet}}]{WalderEtal2012}
{Walder}, R., {Folini}, D., \& {Meynet}, G. 2012, \ssr, 166, 145

\bibitem[{{Wang} {et~al.}(2015){Wang}, {Cui}, {Zhu}, \& {Tian}}]{Wang15}
{Wang}, L., {Cui}, X., {Zhu}, H., \& {Tian}, W. 2015, Advancing Astrophysics
  with the Square Kilometre Array (AASKA14), 64

\bibitem[{{Warren} {et~al.}(2017){Warren}, {Ellison}, {Barkov}, \&
  {Nagataki}}]{WEBN2017}
{Warren}, D.~C., {Ellison}, D.~C., {Barkov}, M.~V., \& {Nagataki}, S. 2017,
  \apj, 835, 248

\bibitem[{{Warren} {et~al.}(2015){Warren}, {Ellison}, {Bykov}, \&
  {Lee}}]{WarrenEllison2015}
{Warren}, D.~C., {Ellison}, D.~C., {Bykov}, A.~M., \& {Lee}, S.-H. 2015,
  \mnras, 452, 431

\bibitem[{{Waxman}(1995)}]{Waxman1995}
{Waxman}, E. 1995, Physical Review Letters, 75, 386

\bibitem[{{Weiler} {et~al.}(2002){Weiler}, {Panagia}, {Montes}, \&
  {Sramek}}]{Weiler02}
{Weiler}, K.~W., {Panagia}, N., {Montes}, M.~J., \& {Sramek}, R.~A. 2002,
  \araa, 40, 387

\bibitem[{{Weiler} {et~al.}(1990){Weiler}, {Panagia}, \& {Sramek}}]{Weiler90}
{Weiler}, K.~W., {Panagia}, N., \& {Sramek}, R.~A. 1990, \apj, 364, 611

\bibitem[{{Weiler} {et~al.}(2011){Weiler}, {Panagia}, {Stockdale}, {Rupen},
  {Sramek}, \& {Williams}}]{Weiler11}
{Weiler}, K.~W., {Panagia}, N., {Stockdale}, C., {et~al.} 2011, \apj, 740, 79

\bibitem[{{Weiler} {et~al.}(1986){Weiler}, {Sramek}, {Panagia}, {van der
  Hulst}, \& {Salvati}}]{Weiler86}
{Weiler}, K.~W., {Sramek}, R.~A., {Panagia}, N., {van der Hulst}, J.~M., \&
  {Salvati}, M. 1986, \apj, 301, 790

\bibitem[{{Weiler} {et~al.}(1991){Weiler}, {van Dyk}, {Discenna}, {Panagia}, \&
  {Sramek}}]{Weiler91}
{Weiler}, K.~W., {van Dyk}, S.~D., {Discenna}, J.~L., {Panagia}, N., \&
  {Sramek}, R.~A. 1991, \apj, 380, 161

\bibitem[{{Weinstein} {et~al.}(2015){Weinstein}, {Aliu}, {Casanova}, {Di
  Girolamo}, {Dyrda}, {Hahn}, {Majumdar}, {Rodriguez}, {Tibaldo}, \& {CTA
  Consortium}}]{WeinsteinEtal2015}
{Weinstein}, A., {Aliu}, E., {Casanova}, S., {et~al.} 2015, ArXiv e-prints

\bibitem[{{Wright} {et~al.}(2015){Wright}, {Drew}, \&
  {Mohr-Smith}}]{Wright2015}
{Wright}, N.~J., {Drew}, J.~E., \& {Mohr-Smith}, M. 2015, \mnras, 449, 741

\bibitem[{{Wright} {et~al.}(2014){Wright}, {Parker}, {Goodwin}, \&
  {Drake}}]{WrightEtal2014}
{Wright}, N.~J., {Parker}, R.~J., {Goodwin}, S.~P., \& {Drake}, J.~J. 2014,
  \mnras, 438, 639

\bibitem[{{Yadav} {et~al.}(2016){Yadav}, {Ray}, \& {Chakraborti}}]{Yadav16}
{Yadav}, N., {Ray}, A., \& {Chakraborti}, S. 2016, \mnras, 459, 595

\bibitem[{{Yoast-Hull} {et~al.}(2017){Yoast-Hull}, {Gallagher}, {Halzen},
  {Kheirandish}, \& {Zweibel}}]{Yoast-Hull2017}
{Yoast-Hull}, T.~M., {Gallagher}, J.~S., {Halzen}, F., {Kheirandish}, A., \&
  {Zweibel}, E.~G. 2017, \prd, 96, 043011

\bibitem[{{Zimbardo} {et~al.}(2015){Zimbardo}, {Amato}, {Bovet}, {Effenberger},
  {Fasoli}, {Fichtner}, {Furno}, {Gustafson}, {Ricci}, \&
  {Perri}}]{Zimbardo2015}
{Zimbardo}, G., {Amato}, E., {Bovet}, A., {et~al.} 2015, Journal of Plasma
  Physics, 81, 495810601

\bibitem[{{Zirakashvili} \& {Ptuskin}(2016)}]{ZP2016}
{Zirakashvili}, V.~N. \& {Ptuskin}, V.~S. 2016, Astroparticle Physics, 78, 28

\end{thebibliography}
\end{document}